\documentclass[jacsat]{achemso}

\setkeys{acs}{
abbreviations=true,
articletitle=true,
biblabel=plain,
chaptertitle=true,
doi=true,
email=true,
etalmode=truncate,
keywords=true,
maxauthors=100,
super=true
}

\setcitestyle{super,open={},close={}}

\usepackage[utf8]{inputenc}
\usepackage[T1]{fontenc}
\usepackage[english]{babel}
\usepackage[version=4]{mhchem}
\usepackage{siunitx}
\usepackage{graphicx}
\usepackage{longtable}

\usepackage{geometry}
\usepackage{caption}
\usepackage{subcaption}
\usepackage{float}
\usepackage{natbib}
\usepackage{setspace}
\usepackage{xkeyval}
\usepackage{array}
\usepackage{booktabs}
\usepackage{listings}
\usepackage{lmodern}
\usepackage{mathpazo}
\usepackage{microtype}
\usepackage{hyperref}
\hypersetup{breaklinks,colorlinks=true,linkcolor=black,filecolor=black,urlcolor=black,citecolor=black}

\usepackage{amsmath}
\usepackage{amssymb}
\usepackage{amsfonts}
\usepackage{latexsym}
\usepackage[compact]{titlesec}
\usepackage[usenames,dvipsnames]{xcolor}
\usepackage{mathptmx}
\usepackage{calc}
\usepackage{titletoc}
\usepackage{lipsum}
\usepackage{multirow}
\usepackage{amstext}
\usepackage{tabularx}
\usepackage{adjustbox}
\usepackage{epsfig,color}
\usepackage[section]{placeins}
\usepackage{natmove}
\usepackage{tablefootnote}
\usepackage{threeparttable}
\usepackage[d]{esvect}
\usepackage{mathtools}
\usepackage{physics}
\usepackage{textcomp}

\usepackage[none]{hyphenat}
\usepackage{xr-hyper}
\vfuzz \hfuzz
\SectionsOn
\SectionNumbersOn
\AbstractOn

\title{Chemical Control of Electronic Structure and Topology in Tellurium-Encapsulated Silicene}

\author{Gabriel Elyas Gama Araújo}
\affiliation{Federal University of Goiás, Institute of Physics, Campus Samambaia, 74960600 Goiânia, Brazil}

\author{André Luis de Oliveira Batista}
\affiliation{Institute of Physics, University of Bras{\'{i}}lia, Bras{\'{i}}lia, 70910-900, Brazil}

\author{Willian Oliveira Santos}
\affiliation[DQ-UFJF]{Group of Physical Chemistry of Solids and Interfaces, Department of Chemistry, Federal University of Juiz de Fora, Juiz de Fora, 36036-330, MG, Brazil}

\author{Alexandre Amaral Leitão}
\affiliation[DQ-UFJF]{Group of Physical Chemistry of Solids and Interfaces, Department of Chemistry, Federal University of Juiz de Fora, Juiz de Fora, 36036-330, MG, Brazil}

\author{Alexandre Cavalheiro Dias}
\affiliation{Institute of Physics, University of Bras{\'{i}}lia, Bras{\'{i}}lia, 70910-900, Brazil}
\alsoaffiliation{International Center of Physics, University of Bras{\'{i}}lia, Bras{\'{i}}lia, 70910-900, Brazil}
\alsoaffiliation{Computational Materials Laboratory, LCCMat, Institute of Physics, University of Bras{\'{i}}lia, $70910-900$, Bras{\'{i}}lia, Brazil}

\author{Andreia Luisa da Rosa}
\email{andreialuisa@ufg.br}
\affiliation{Federal University of Goiás, Institute of Physics, Campus Samambaia, 74960600 Goiânia, Brazil}


\date{\today}

\begin{document}

\maketitle

\begin{abstract}

We investigate the structural, vibrational, thermodynamic, electronic, and
optical properties of two-dimensional \ce{Si2X2Te2} monolayers
(\ce{X}= \ce{B}, \ce{Al}, \ce{Ga}, and \ce{In}) using first-principles calculations. The optimized structures preserve a common \ce{Si}--\ce{X}--\ce{Te} framework, while their lattice parameters and vibrational spectra evolve systematically with chemical substitution at the group-III site. Phonon calculations show no imaginary frequencies along the investigated high-symmetry paths, supporting the harmonic dynamical stability of all four monolayers. \ce{Si2B2Te2} exhibits the highest optical phonon frequencies, reflecting the combined effects of the low \ce{B} atomic mass and its local bonding environment, whereas the \ce{Ga}- and \ce{In}-containing systems display progressively softer vibrational modes.

Electronic-structure calculations performed with the PBE and HSE06
functionals, including spin--orbit coupling, show that all compounds are
semiconductors. The valence-band edge is predominantly derived from \ce{Te}
$p$ states, whereas the conduction-band edge has a mixed \ce{Si}--\ce{X}--\ce{Te} $p$-orbital character. Spin--orbit coupling modifies the dispersion near
avoided crossings and regions of strong orbital hybridization, with the
largest relativistic effects occurring in the \ce{In}-containing monolayer.

The in-plane dielectric response is evaluated within the
independent-particle approximation and by solving the Bethe--Salpeter
equation using a Wannier-based excitonic Hamiltonian. Electron--hole
interactions produce modest shifts and redistribution of oscillator strength 
near the optical onset, while the response remains weakly anisotropic within
the monolayer plane. The calculated $\mathbb{Z}_2$ invariants classify
\ce{Si2B2Te2}, \ce{Si2Al2Te2}, and \ce{Si2Ga2Te2} as
topologically trivial, whereas \ce{Si2In2Te2} is identified as a
candidate quantum spin Hall insulator. These results demonstrate that
chemical substitution at the group-III site provides an effective route for
tuning the vibrational, electronic, optical, and topological properties of
\ce{Si2X2Te2} monolayers.

\end{abstract}

\section{Introduction}

The discovery of graphene has triggered an intense search for novel two-dimensional (2D) materials with tunable electronic, optical, and topological properties.\cite{Novoselov2004,Geim2007} Beyond graphene, layered systems such as transition metal dichalcogenides (TMDs), group-IV analogues, and elemental tellurium allotropes have demonstrated a rich variety of physical phenomena, including strong spin–orbit coupling (SOC), reduced dielectric screening, and enhanced many-body interactions.\cite{Xu2014,TellureneReview} In particular, the reduced dimensionality of 2D materials leads to pronounced excitonic effects and enables the emergence of nontrivial topological phases, making them attractive platforms for next-generation optoelectronic and quantum devices.\cite{QSHKaneMele,Bernevig2006}

A central theme in contemporary condensed matter physics is the interplay between electronic band topology and many-body interactions.\cite{HasanKane2010,QiZhang2011} Quantum spin Hall (QSH) insulators, characterized by an insulating bulk and topologically protected helical edge states, arise from SOC-induced band inversion and are robust against nonmagnetic perturbations.\cite{QSHKaneMele,Bernevig2006} While several 2D topological insulators have been predicted and experimentally realized, the search for new material platforms with sizable band gaps and tunable properties remains ongoing.

Tellurium-based systems have recently attracted significant attention due to their strong SOC originating from heavy $p$ orbitals and their structural versatility in low dimensions.\cite{TellureneReview} Monolayer and few-layer tellurene exhibit diverse allotropes (e.g., $\alpha$, $\beta$, and square phases) with tunable electronic properties, including semiconducting, metallic, and topological regimes.\cite{TellureneAlphaBeta} The presence of strong SOC combined with reduced screening in 2D tellurium systems also leads to large exciton binding energies, making them ideal candidates for studying the interplay between topology and excitonic effects.\cite{TellureneReview}

In parallel, chemical functionalization and compositional engineering have emerged as powerful strategies to tailor the electronic structure of 2D materials.\cite{Chhowalla2016} In particular, ternary and quaternary compounds incorporating group-IV elements (such as \ce{Si}) and group-III elements (such as \ce{B}, \ce{Al}, and \ce{Ga}) offer additional degrees of freedom to control band alignment, symmetry, and SOC strength. For instance, breaking inversion symmetry through asymmetric functionalization can induce Rashba-type spin splitting, while the incorporation of heavier elements can enhance SOC-driven band inversion.\cite{Rashba1960,Manchon2015}

Motivated by these considerations, we investigate a class of largely unexplored 2D compounds with chemical composition \ce{Si2X2Te2} (\ce{X}= \ce{B}, \ce{Al}, \ce{Ga}, \ce{In}). These materials combine a \ce{Si}–\ce{Te} backbone with group-III elements that act as electronic and structural modifiers, potentially enabling tunable band gaps, symmetry breaking, and SOC effects. From a chemical perspective, the choice of \ce{X} provides a systematic route to tune the electronic structure: lighter elements such as \ce{B} tend to open larger band gaps, while heavier elements such as In can enhance relativistic effects and promote stronger SOC-induced modifications of the electronic structure.

In this work, we first analyze the electronic band structures of \ce{Si2X2Te2} monolayers using first-principles calculations. The presence of \ce{Te}-derived $p$ states near the band edges suggests that SOC may play a crucial role in shaping the low-energy electronic properties. This raises the possibility that these systems could host nontrivial topological phases, depending on the details of orbital hybridization and crystal symmetry. Furthermore, the reduced dimensionality and the expected moderate-to-large band gaps make these materials promising candidates for exhibiting strong excitonic effects, which could be further influenced by the underlying band topology.

Our study aims to establish \ce{Si2X2Te2} as a versatile platform for exploring the interplay between chemical composition, spin–orbit coupling, and emergent quantum phases in two dimensions. By providing a detailed analysis of their electronic structure, we lay the groundwork for future investigations of their topological properties and excitonic behavior.

\section{Computational Methodology}

First-principles calculations were performed within density functional theory (DFT) using the Vienna \textit{Ab initio} Simulation Package (VASP)~\cite{Kresse1996CMS,Kresse1996PRB,Kresse1999PAW}. The interaction between valence electrons and ionic cores was described using the projector augmented-wave (PAW) method~\cite{Blochl_17953_1994}, while exchange--correlation effects were treated within the generalized-gradient approximation of Perdew, Burke, and Ernzerhof (PBE)~\cite{Perdew_3865_1996}. A plane-wave kinetic-energy cutoff of $2\times\mathrm{ENMAX}_{\max}$ was employed for structural optimization, where $\mathrm{ENMAX}_{\max}$ is the largest recommended cutoff energy among the PAW projectors used. For the remaining calculations, a cutoff of $1.125\times\mathrm{ENMAX}_{\max}$ was adopted. Details of the PAW projectors are provided in the Supporting Information (SI). The Brillouin zone was sampled using a $\Gamma$-centered $12\times12\times1$ Monkhorst--Pack mesh, and all structures were fully relaxed until the residual forces on each atom were smaller than \SI{1E-2}{\electronvolt/\angstrom}. A vacuum region of \SI{28}{\angstrom}  was introduced along the out-of-plane direction to eliminate interactions between periodic images.

Spin--orbit coupling (SOC) was included self-consistently in all electronic-structure calculations. To obtain a more accurate description of the electronic properties, hybrid-functional calculations were additionally performed using the HSE06 screened hybrid functional~\cite{Heyd_219906_2006}.

The vibrational properties were calculated using the finite-displacement method as implemented in the PHONOPY package~\cite{Togo2015Phonopy,Togo2023Phonopy}, considering a $2\times2\times1$ supercell with the same \textbf{k}-points density of previous calculations. The absence of imaginary phonon frequencies confirmed the harmonic dynamical stability of all investigated monolayers. Thermodynamic quantities, including the Helmholtz free energy, entropy, and constant-volume heat capacity, were obtained within the harmonic approximation from the calculated phonon spectra.

Maximally localized Wannier functions (MLWFs) were generated using the Wannier90 package~\cite{Pizzi2020Wannier90}. The resulting Wannier Hamiltonians accurately reproduce the HSE06+SOC electronic bands near the Fermi level and were employed as the basis for the optical and topological calculations.

The linear optical response was evaluated within both the independent-particle approximation (IPA) and the Bethe--Salpeter equation (BSE) formalism using the WANTIBEXOS package~\cite{Dias2023WanTiBEXOS}. The BSE Hamiltonian is constructed using the MLWF tight binding Hamiltonian to describe the single-particle states, with a \textbf{k}-points density of \SI{120}{\per\angstrom} in  the xy plane, considering the following number of conduction $n_c$ and valence $n_v$ states: $n_{c}=9,n_{v}=9$ for \ce{Si2B2Te2}, $n_{c}=10,n_{v}=10$ for \ce{Si2Al2Te2}, $n_{c}=9,n_{v}=12$ for \ce{Si2Ga2Te2} and $n_{c}=10,n_{v}=12$ for \ce{Si2In2Te2}, which are sufficient to describe the optical response. The electron-hole Coulomb potential, in BSE, was modeled by a 2D Coulomb truncated potential (V2DT) \cite{Rozzi_205119_2006}, a smearing of \SI{0.05}{\electronvolt} was used to describe the real and imaginary parts of the dielectric function. Optical spectra were calculated for light polarized along the two in-plane crystallographic directions.

The topological properties were investigated using the Wannier-interpolated Hamiltonians. Berry curvature was calculated from the Kubo formalism, while the $\mathbb{Z}_2$ topological invariant was determined from the evolution of the hybrid Wannier charge centers (Wilson loops) using WannierTools~\cite{Wu2018WannierTools}. This approach enables an unambiguous classification of the electronic topology of the investigated monolayers.

\section{Results and discussion}

\subsection{Structural Properties}

\begin{figure}[!htb]
\centering
\includegraphics[width=\columnwidth,clip=true,keepaspectratio]{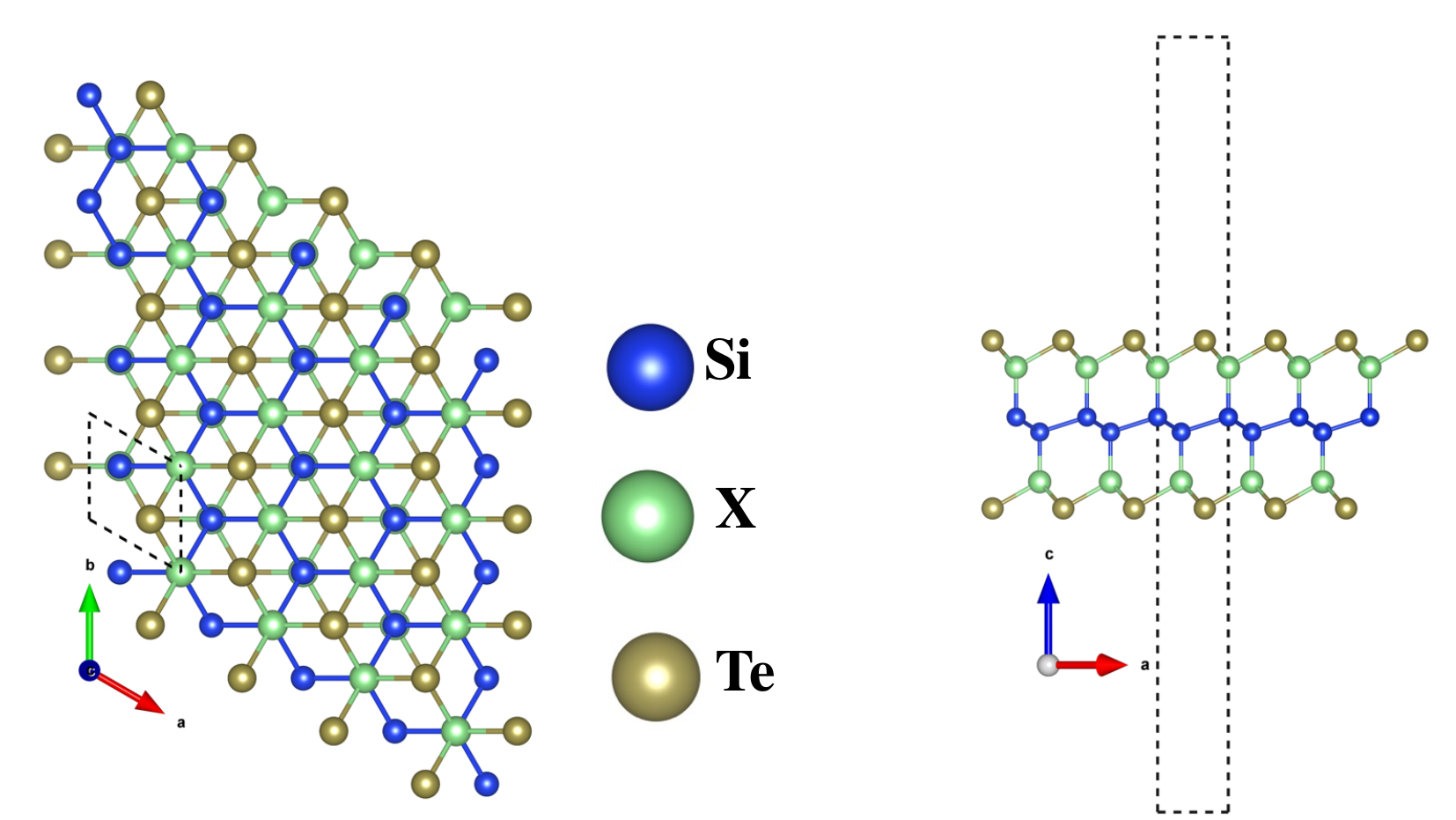}
\caption{Top and side views of the optimized \ce{Si2X2Te2}
(\ce{X}= \ce{B}, \ce{Al}, \ce{Ga}, and \ce{In}) monolayer structure. Blue, green, and gold spheres represent \ce{Si}, \ce{X}, and \ce{Te} atoms, respectively. The in-plane lattice vectors are indicated by $\mathbf{a}$ and $\mathbf{b}$.}
\label{fig:all_relaxed}
\end{figure}

The optimized \ce{Si2X2Te2} (\ce{X}= \ce{B}, \ce{Al}, \ce{Ga}, and \ce{In}) monolayers share the same structural framework, as illustrated in Fig.~\ref{fig:all_relaxed}. The structure consists of a two-dimensional \ce{Si}--\ce{X}--\ce{Te} network, with the group-III element occupying equivalent sites within the monolayer. The optimized geometries exhibit a non-planar atomic arrangement, reflecting the different local coordination environments of \ce{Si}, \ce{X}, and \ce{Te}.

A systematic expansion of the lattice is observed upon substitution
of \ce{B} by the heavier group-III elements. The optimized \ce{Si2X2Te2} monolayers preserve the same two-dimensional lattice geometry across the series. The calculated in-plane lattice constants are \SI{3.678}{}, \SI{4.022}{}, \SI{4.026}{}, and \SI{4.151}{\angstrom} for \ce{X} = \ce{B}, \ce{Al}, \ce{Ga}, and \ce{In}, respectively. The progressive lattice expansion from \ce{B} to \ce{In} reflects the increasing atomic size of the group-III element, while the overall structural framework remains unchanged. The \ce{Al}- and \ce{Ga}-containing monolayers therefore exhibit nearly identical equilibrium lattice parameters, indicating that substitution of \ce{Al} by \ce{Ga} produces only minor changes in the in-plane dimensions.

In contrast, the smaller \ce{B} atom leads to a more compact lattice, consistent with the shorter characteristic length scales expected for the \ce{B}-containing structure. Substitution by \ce{In} produces a further expansion of the lattice, following the increase in atomic size along the group-III series. The structural evolution across the \ce{Si2X2Te2} family therefore provides the first indication that chemical substitution at the \ce{X} site can systematically modify the local bonding environment and, consequently, the vibrational and electronic properties discussed below.

\subsection{Phonon Dispersion and Dynamical Stability}

\begin{figure}[H]
\centering
\includegraphics[width=1\linewidth]{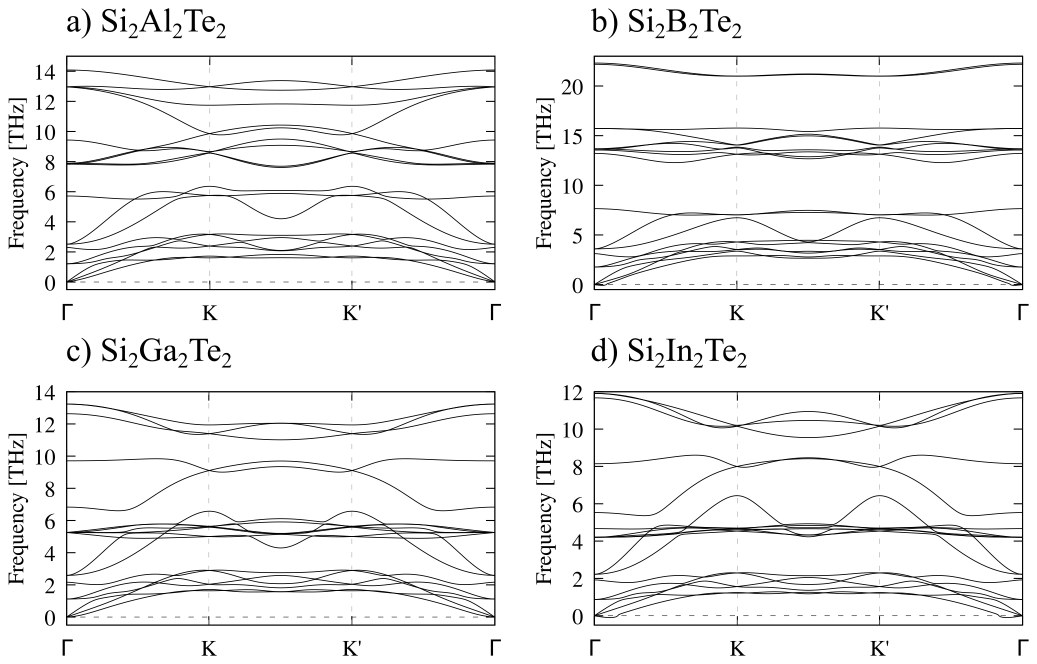}
\caption{Phonon dispersion relations calculated along the high-symmetry paths of the Brillouin zone for (a) \ce{Si2Al2Te2} , (b) \ce{Si2B2Te2}, (c) \ce{Si2Ga2Te2} and (d) \ce{Si2In2Te2}. }
\label{fig:phonons}
\end{figure}

The phonon dispersions of \ce{Si2X2Te2} (\ce{X}= \ce{B}, \ce{Al}, \ce{Ga}, and \ce{In}) along the selected high-symmetry paths of the Brillouin zone are shown in Fig.~\ref{fig:phonons}. The lack of imaginary phonon frequencies observed along these paths supports the harmonic dynamical stability of all four monolayers. The acoustic branches approach zero frequency at $\Gamma$, as expected for a free-standing two-dimensional crystal. In particular, the lowest acoustic branch displays the characteristic soft dispersion associated with the out-of-plane flexural ZA mode.

A clear chemical dependence is observed in the optical phonon branches. \ce{Si2B2Te2} exhibits the highest maximum phonon frequency, reaching approximately \SI{22}{}--\SI{23}{THz}, whereas the \ce{Al}-, \ce{Ga}-, and \ce{In}-containing monolayers display substantially lower maximum
frequencies, in the range of approximately \SI{12}{}--\SI{14}{THz}. The pronounced shift toward higher frequencies in the \ce{B}-containing system reflects primarily the much lower atomic mass of \ce{B}, together with differences in the local force constants associated with the \ce{B}-containing bonding environment.

Moving from \ce{B} to the heavier group-III elements leads to a progressive
shift of several vibrational branches toward lower frequencies. This
trend is consistent with the approximate mass dependence
$\omega \propto \sqrt{\frac{k}{M}}$, where $k$ represents an effective force constant and $M$ an effective vibrating mass. The phonon-frequency evolution therefore reflects the combined effects of increasing atomic mass and changes in the local bonding environment across the \ce{Si2X2Te2} series.

The \ce{Al}- and \ce{Ga}-based compounds display rather similar phonon spectra, consistent with their closely related optimized lattice parameters, whereas \ce{Si2In2Te2} exhibits a further softening of several
optical branches. These results indicate that chemical substitution at
the group-III site provides an effective mechanism for tuning the
vibrational energy scale of the monolayers.

\begin{figure}[!hbt]
\centering
\includegraphics[width=1\linewidth,clip=true]{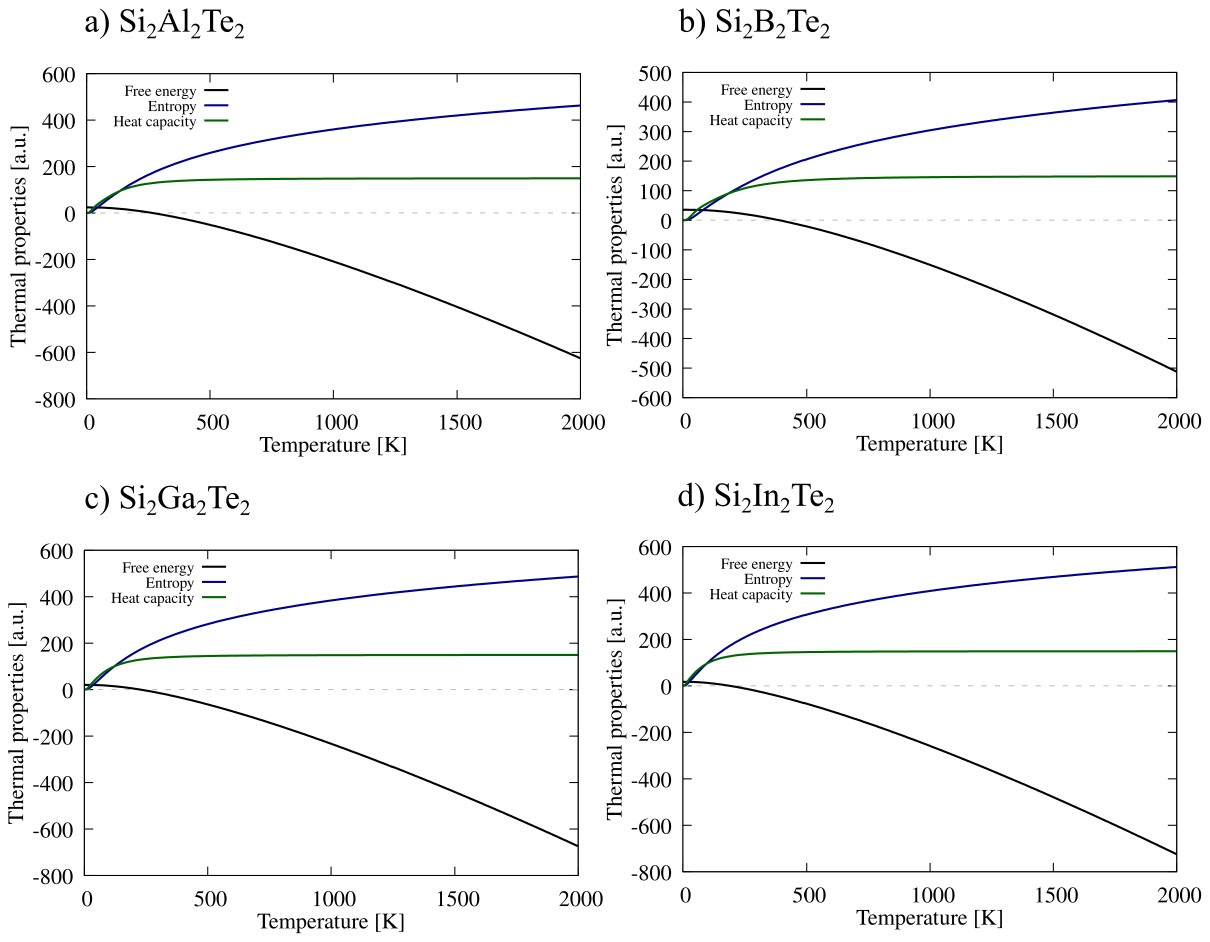}
\caption{Temperature dependence of the vibrational thermodynamic properties of \ce{Si2X2Te2} (\ce{X}= \ce{B}, \ce{Al}, \ce{Ga}, and \ce{In}) obtained within the harmonic approximation: (a) vibrational Helmholtz free energy $F_{\rm vib}$, (b) vibrational entropy $S_{\rm vib}$, and (c) constant-volume heat capacity $C_V$.}
\label{fig:thermal}
\end{figure}

The temperature dependence of the vibrational thermodynamic properties
of the \ce{Si2X2Te2} monolayers is shown in Fig.~\ref{fig:thermal}. The Helmholtz free energy, vibrational entropy, and constant-volume heat capacity were obtained from the calculated phonon spectra within the harmonic approximation.

For all four compounds, the vibrational contribution to the Helmholtz free energy decreases monotonically with increasing temperature, whereas the entropy increases as a progressively larger number of phonon modes becomes thermally populated. The corresponding heat capacity increases rapidly in the low-temperature regime and gradually approaches a nearly constant value at high temperatures, consistent with the classical high-temperature limit of the lattice contribution to the heat capacity.

Differences among the compounds are most pronounced at low and intermediate temperatures and reflect the different phonon-frequency distributions discussed above. In particular, the higher-frequency modes of \ce{Si2B2Te2} require larger thermal energies to become fully populated, whereas the softer vibrational modes of the \ce{Al}-, \ce{Ga}-, and \ce{In}-containing systems contribute at comparatively lower temperatures. Chemical substitution at the group-III site therefore modifies not only the phonon spectrum but also the temperature scale over which the vibrational degrees of freedom become thermally activated.

It should be emphasized that these quantities were obtained within the harmonic approximation. Consequently, the high-temperature behavior represents the harmonic vibrational contribution and does not, by itself, establish structural or thermal stability at such temperatures, where  anharmonic effects and possible structural transformations may become important.

\subsection{Electronic Structure and Orbital Character}

The electronic structures of the \ce{Si2X2Te2} (\ce{X}= \ce{B}, \ce{Al}, \ce{Ga}, and \ce{In}) monolayers were first calculated at the PBE and PBE+SOC levels. The corresponding band structures and projected densities of states are presented in the Supporting Information (SI). All four
monolayers exhibit semiconducting behavior with indirect band gaps.

\begin{figure}[H]
\centering
\begin{subfigure}{0.68\textwidth}
\subcaption[]{}
\includegraphics[width=\textwidth,clip=true]{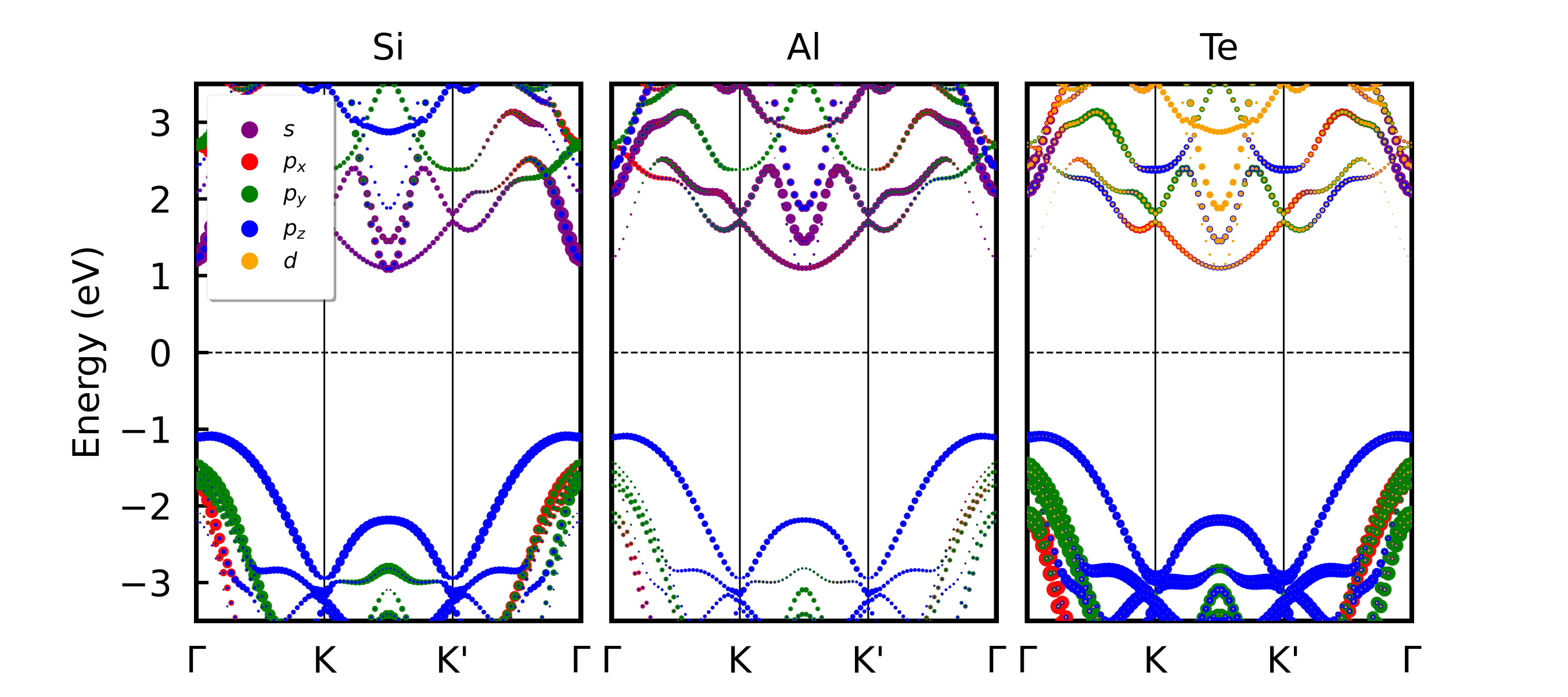}
\end{subfigure}
\hfill
\begin{subfigure}{0.68\textwidth}
\subcaption[]{}
\includegraphics[width=\textwidth,clip=true]{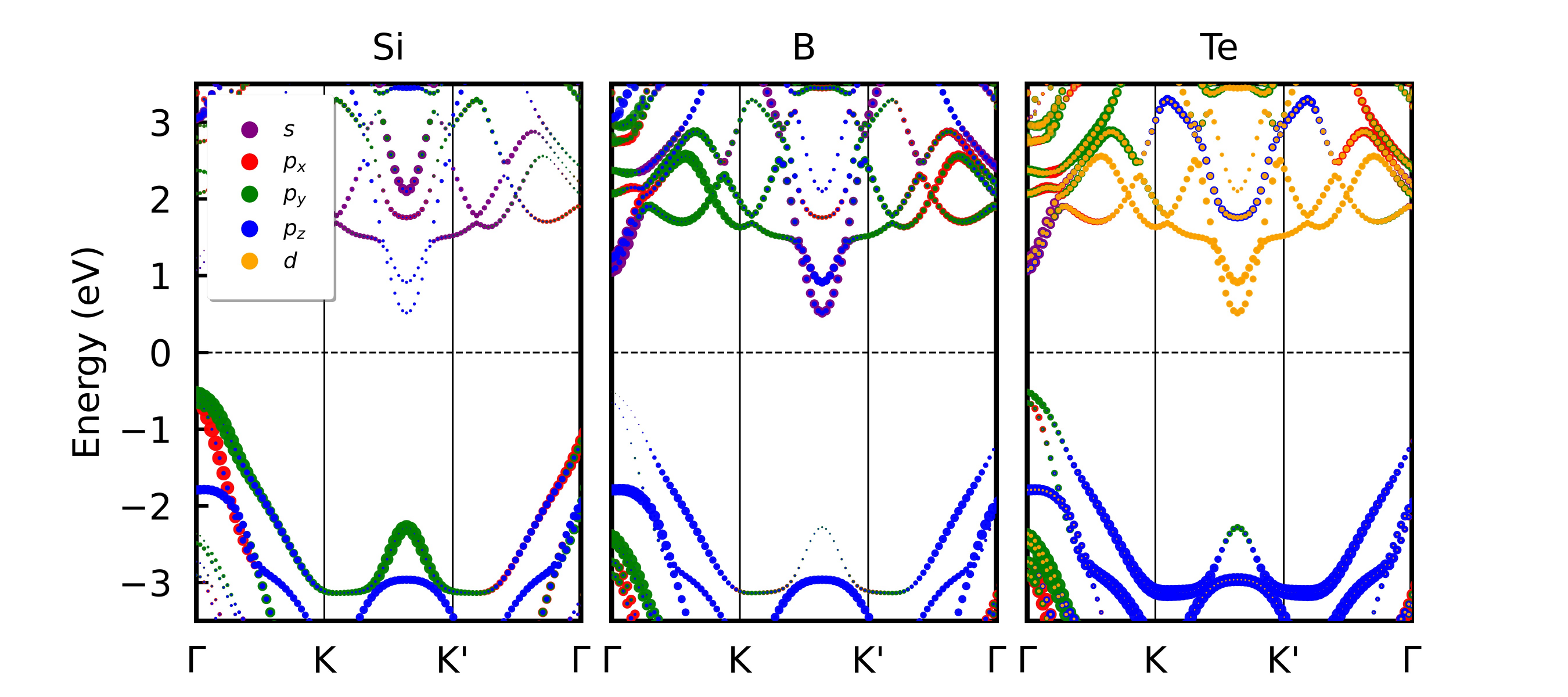}
\end{subfigure}
\hfill
\begin{subfigure}{0.68\textwidth}
\subcaption[]{}
\includegraphics[width=\textwidth,clip=true]{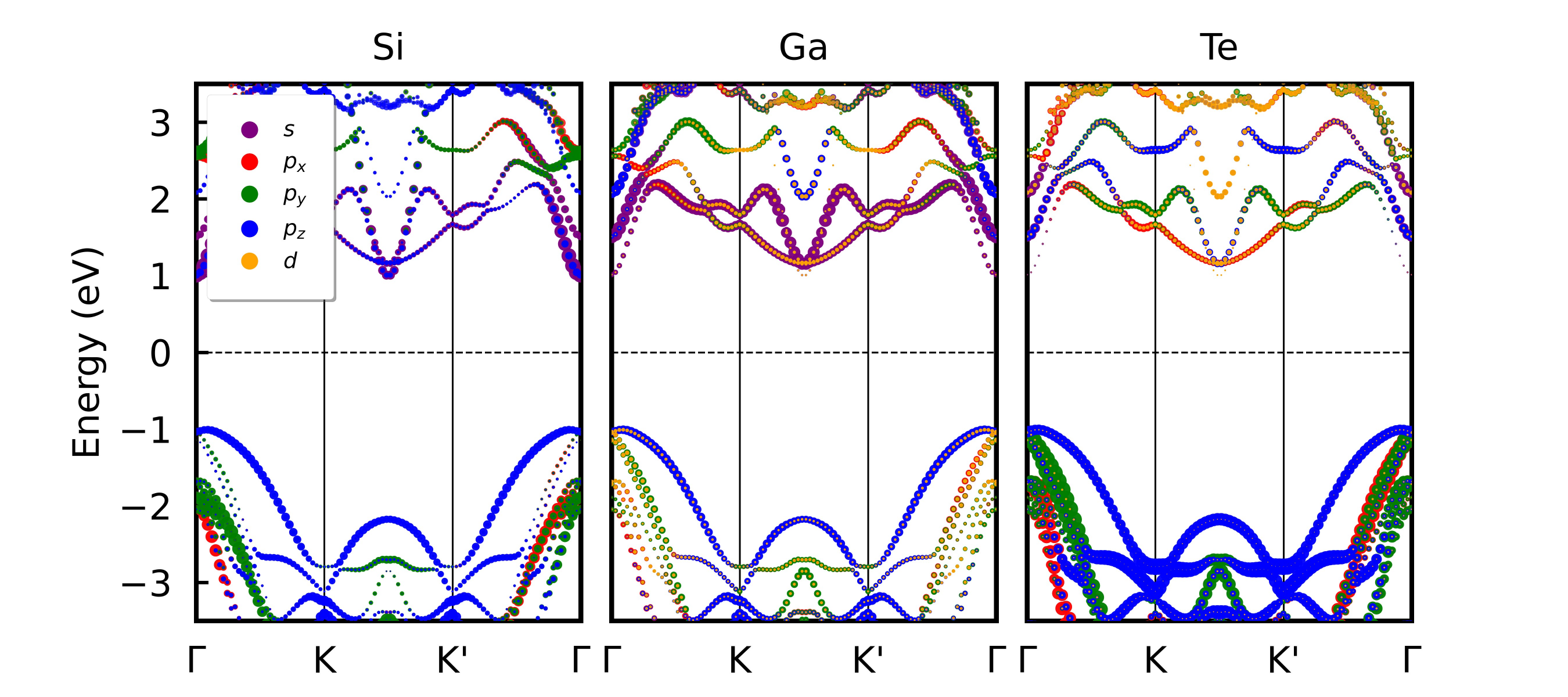}
\end{subfigure}
\hfill
\begin{subfigure}{0.68\textwidth}
\subcaption[]{}
\includegraphics[width=\textwidth,clip=true]{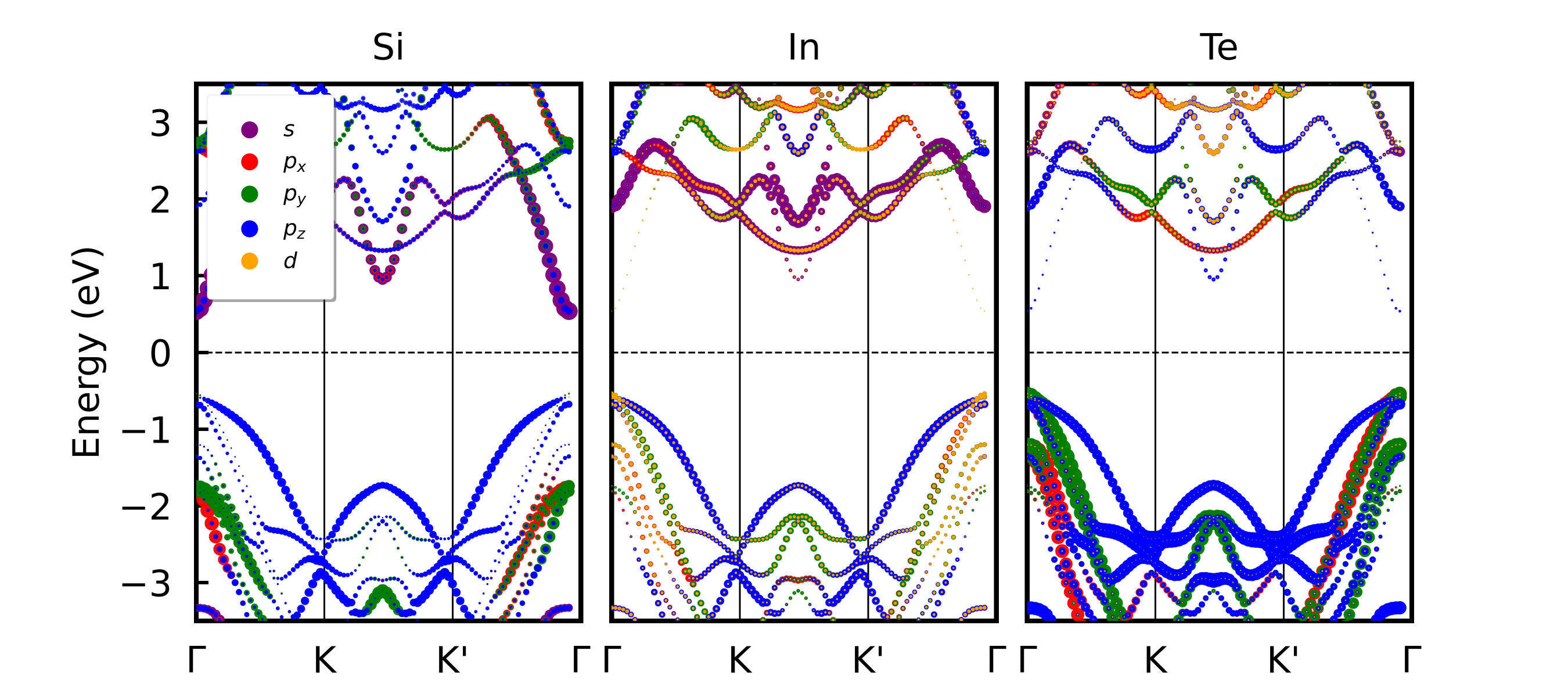}
\end{subfigure}
\caption{\label{fig:bands_proj_hse} Electronic band structure calculated with HSE06+SOC of (a) \ce{Si2Al2Te2}, (b) \ce{Si2B2Te2}, (c) \ce{Si2Ga2Te2}, and (d) \ce{Si2In2Te2}.}
\end{figure}

The inclusion of spin--orbit coupling modifies the electronic dispersion, particularly in regions containing closely spaced or nearly degenerate bands. These effects are especially evident in the valence-band manifold, which contains a substantial contribution from the \ce{Te} $5p$ states. Nevertheless, SOC does not qualitatively alter the overall electronic structure at the PBE level.

Because semi-local functionals generally underestimate semiconductor band gaps, the electronic structures were further evaluated using the HSE06 \cite{Heyd_219906_2006} hybrid functional including SOC. A comparison between the PBE+SOC and HSE06+SOC results is provided in the Supporting Information. HSE06+SOC increases the separation between the valence and conduction manifolds while largely preserving their overall dispersion and ordering. This indicates that the qualitative features of the electronic structure are robust with respect to the choice of exchange--correlation functional.

The orbital-projected HSE06+SOC band structures are shown in Fig.~\ref{fig:bands_proj_hse}. The states near the band edges are predominantly of $p$ character, although their atomic contributions differs considerably between the valence and conduction manifolds. The upper valence bands are dominated by \ce{Te} $5p$ states, with smaller contributions from \ce{Si} and the group-III element. In contrast, the lowest conduction bands exhibit a more strongly mixed orbital character involving \ce{Si}, \ce{X}, and \ce{Te} $p$ states. The electronic transitions near the fundamental gap therefore involve predominantly \ce{Te}-derived valence states and conduction states with mixed \ce{Si}--\ce{X}--\ce{Te} character.

The orbital composition evolves systematically with substitution at the \ce{X} site. In \ce{Si2B2Te2}, the \ce{B}-derived states contribute differently to the frontier bands compared with the heavier group-III elements, consistent with the distinct electronic dispersion observed for this compound. The \ce{Al}- and \ce{Ga}-containing systems display qualitatively similar band structures and orbital compositions, in agreement with their closely related equilibrium structures.

For \ce{Si2In2Te2}, the \ce{In}-derived $p$ states contribute appreciably to the conduction-band manifold. Together with the strong \ce{Te} contribution to the valence bands, the presence of the heavier \ce{In} atom enhances the relevance of relativistic effects in the electronic structure. This compound therefore deserves particular attention in the analysis of the topological properties discussed below.

The orbital projections also reveals distinct contributions from the $p_x$, $p_y$, and $p_z$ components across the valence and conduction bands. This orbital dependence provides a microscopic basis for understanding the polarization dependence of the optical response.

A direct comparison between the PBE band structures calculated without and
with spin--orbit coupling provides additional insight into the role of SOC
across the $\mathrm{Si_2X_2Te_2}$ series. As shown in Figs.~S1 and S2,
the inclusion of SOC preserves the overall semiconducting character of all
four monolayers but modifies the dispersion of several bands near the band
edges, particularly in regions where multiple $p$-derived states are closely
spaced in energy.

For $\mathrm{Si_2B_2Te_2}$, $\mathrm{Si_2Al_2Te_2}$, and
$\mathrm{Si_2Ga_2Te_2}$, the SOC-induced changes remain comparatively modest
and do not lead to an evident qualitative rearrangement of the frontier-band
manifold. In contrast, $\mathrm{Si_2In_2Te_2}$ exhibits a noticeably stronger
SOC-induced modification of the low-energy electronic structure, especially
in the conduction-band region. This enhanced sensitivity is consistent with
the larger relativistic contribution introduced by the In-containing
environment and with the appreciable participation of In-$p$ states in the
conduction manifold.

Importantly, the orbital-projected bands reveal that the states close to the
fundamental gap are already strongly hybridized, involving mainly Te-$p$
states in the valence region and mixed Si--X--Te $p$ character in the
conduction region. Therefore, the present band structures do not provide an
unambiguous signature of a simple exchange of two well-defined orbital
characters upon inclusion of SOC. Rather, the results indicate an
SOC-induced reorganization of a strongly hybridized $p$-orbital manifold,
which is substantially more pronounced in $\mathrm{Si_2In_2Te_2}$ than in
the topologically trivial members of the series.

This distinction is relevant for the interpretation of the topological phase.
The nontrivial character of $\mathrm{Si_2In_2Te_2}$ should therefore be
associated with the action of SOC on the chemically hybridized Si--In--Te
electronic states, while the topological classification itself is established
independently from the evolution of the Wannier charge centers and the
corresponding $Z_2$ invariant.
\subsection{Charge-density redistribution and electronic coupling}

To further characterize the electronic coupling between the Si sublattice and
the surrounding X--Te framework, we evaluated the charge-density difference,
defined as

\begin{equation}
\Delta\rho(\mathbf{r}) =
\rho_{\mathrm{Si_2X_2Te_2}}(\mathbf{r})
-\rho_{\mathrm{Si_2}}(\mathbf{r})
-\rho_{\mathrm{X_2Te_2}}(\mathbf{r}),
\end{equation}

where the charge densities of the isolated $\mathrm{Si_2}$ and
$\mathrm{X_2Te_2}$ fragments were calculated in the same unit cell and with
the atomic positions fixed at those of the fully relaxed
$\mathrm{Si_2X_2Te_2}$ monolayer. With this construction, positive and
negative values of $\Delta\rho(\mathbf{r})$ correspond, respectively, to
electron accumulation and depletion induced by the interaction between the
two fragments.

\begin{figure}[H]
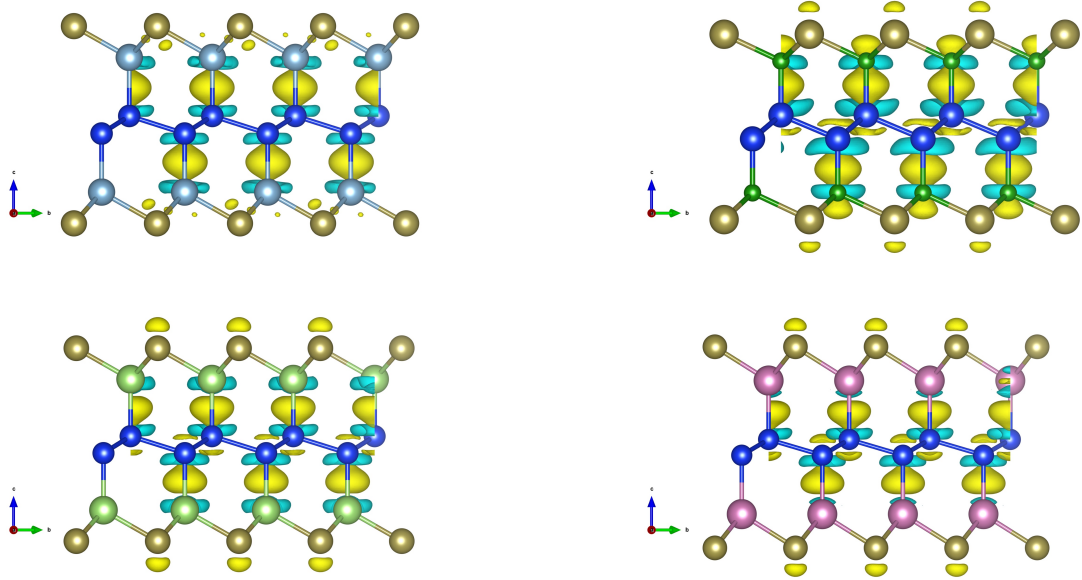

\centering
\begin{subfigure}{0.47\textwidth}
\includegraphics[width=\textwidth,clip=true]{CDD_SiAlTe.pdf}
\label{fig:cdd_al}
\end{subfigure}
\hfill
\begin{subfigure}{0.47\textwidth}
\includegraphics[width=\textwidth,clip=true]{CDD_SiBTe.pdf}
\label{fig:cdd_b}
\end{subfigure}
\hfill
\begin{subfigure}{0.47\textwidth}
\includegraphics[width=\textwidth,clip=true]{CDD_SiGaTe.pdf}
\label{fig:cdd_ga}
\end{subfigure}
\hfill
\begin{subfigure}{0.47\textwidth}
\includegraphics[width=\textwidth,clip=true]{CDD_SiInTe.pdf}
\label{fig:cdd_in}
\end{subfigure}
\caption{Charge density difference ($\Delta\rho$) for the monolayers of (a) $\text{Si}_2\text{Al}_2\text{Te}_2$, (b) $\text{Si}_2\text{B}_2\text{Te}_2$, (c) $\text{Si}_2\text{Ga}_2\text{Te}_2$, and (d) $\text{Si}_2\text{In}_2\text{Te}_2$. The yellow and blue isosurfaces represent regions of electron accumulation ($\Delta\rho > 0$) and depletion ($\Delta\rho < 0$), respectively. The isosurface level is set to $0.003 \, e/\text{\AA}^3$.}
\label{fig:charge_density_difference}
\end{figure}

Figure~\ref{fig:charge_density_difference} shows the resulting charge-density
difference for the four members of the $\mathrm{Si_2X_2Te_2}$ family.
In all cases, the redistribution of charge is concentrated not only around
the atomic sites but also in the regions connecting the Si sublattice to the
neighboring X--Te units. This spatial pattern indicates that the electronic
structure of the complete monolayer cannot be described as a simple
superposition of weakly interacting $\mathrm{Si_2}$ and $\mathrm{X_2Te_2}$
subsystems. Instead, the charge redistribution is consistent with appreciable
chemical coupling and hybridization between Si-derived and X/Te-derived
states.

The evolution of $\Delta\rho(\mathbf{r})$ across the series also reflects the
modification of this coupling upon chemical substitution at the group-III
site. In particular, comparison between $\mathrm{Si_2Ga_2Te_2}$ and
$\mathrm{Si_2In_2Te_2}$ is relevant because these systems retain closely
related structural and electronic features while exhibiting different
topological classifications. The corresponding charge-density-difference
maps display changes in the spatial distribution of the accumulation and
depletion regions around the Si--X--Te network, indicating that substitution
of Ga by In modifies the electronic environment experienced by the Si
sublattice.

It should be emphasized that the charge-density-difference isosurfaces provide
a qualitative picture of electronic redistribution and should not be
interpreted as a direct measure of charge transfer or as evidence of band
inversion. In particular, the nontrivial $Z_2$ invariant of
$\mathrm{Si_2In_2Te_2}$ cannot be inferred from $\Delta\rho(\mathbf{r})$
alone. Rather, the charge-density analysis complements the orbital-resolved
band structures, which show predominantly Te-$p$ character at the valence-band
edge and a strongly hybridized Si--X--Te $p$ manifold in the conduction
region.

Taken together with the SOC-dependent electronic structure and the evolution
of the Wannier charge centers, these results support a picture in which the
nontrivial phase of $\mathrm{Si_2In_2Te_2}$ emerges from the interplay
between chemical hybridization within the Si--In--Te network and spin--orbit
coupling. The charge-density redistribution therefore provides a microscopic
view of the electronic coupling underlying the low-energy band structure,
while the topological character itself is established independently by the
$Z_2$ analysis.

\subsection{Optical Properties}

\begin{figure}[!hbt]
\centering
\includegraphics[width=1\linewidth,clip=true]{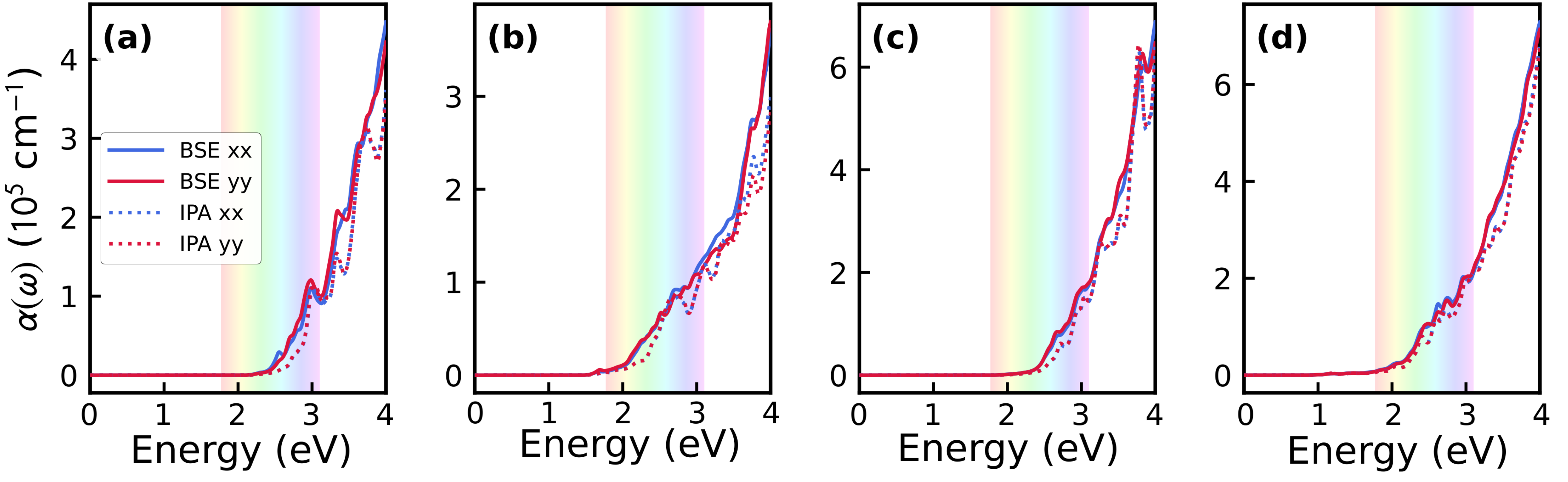}
\caption{Calculated absorption coefficient $\alpha (\omega)$  (a) \ce{Si2Al2Te2}, (b) \ce{Si2B2Te2}, (c) \ce{Si2Ga2Te2} and (d) \ce{Si2In2Te2}. Results are presented within the independent-particle approximation (IPA, dashed lines) and by solving the Bethe-Salpeter equation (BSE, solid lines). Blue and red curves represent light polarization along the $x$ (blue curves) and $y$ (red curves) crystallographic directions, respectively.}
\label{fig:optical}
\end{figure}

\begin{table}
\centering
\caption{Excitonic properties obtained by MLWF-TB+BSE at HSE06: fundamental band gap ($E_{g}$), direct band gap ($E^{d}_{g}$), direct exciton ground state ($Ex^{d}_{gs}$). The exciton binding energy ($Ex_{b}$) is calculated as $E^{d}_{g}-Ex^{d}_{gs}$. Electronic bandgap (in eV) and exciton binding energies (in meV) of selected two-dimensional materials are also shown.}
\begin{tabular}{lcccc} \toprule
Structure & $E_{g}$ (\si{\electronvolt}) & $E^{d}_{g}$ (\si{\electronvolt})  & $Ex^{d}_{gs}$ (\si{\electronvolt}) & $Ex_{b}$ (\si{\milli\electronvolt}) \\ \midrule
\ce{Si2Al2Te2} &  2.14 (i) &  2.29  & 2.25   &  39.83\\ 
\ce{Si2B2Te2}  &  1.02 (i) &  1.64  & 1.62   &  16.58\\
\ce{Si2Ga2Te2} &  1.96 (i) &  2.00  & 1.96   &  50.96\\ 
\ce{Si2In2Te2} &  1.02 (d) &  1.02  & 1.01   &  5.59\\ 
\bottomrule
\end{tabular}
\label{tab:exciton_data}
\end{table}

A comparison of the optical absorption spectra obtained within the independent-particle approximation (IPA) and including electron–hole interactions through the Bethe–Salpeter equation (BSE) is shown in Fig.~\ref{fig:optical}. The spectra are shown for light polarized along the two in-plane crystallographic directions ($xx$ and $yy$), allowing the effects of excitonic correlations and optical anisotropy to be evaluated.

The inclusion of electron--hole interactions leads to a systematic enhancement of the absorption coefficient near the optical onset compared with the IPA results. The BSE spectra exhibit a slight redshift of the first absorption features together with a redistribution of oscillator strength toward lower photon energies, which is the expected signature of excitonic effects in two-dimensional semiconductors. The relatively small differences between the IPA and BSE spectra indicate that excitonic effects are moderate in the \ce{Si2X2Te2} monolayers, suggesting that the electron–hole interaction modifies the optical response without dramatically changing the overall spectral profile. This behavior is consistent with the relatively small exciton binding energies, which range from approximately 5 to 51 meV, as shown in Table 1. The first optical transitions originate primarily from excitations between Te-derived valence states and hybridized Si/X p conduction states. These monolayers absorb in the visible and ultraviolet spectral regions.

The calculated spectra also reveal weak in-plane optical anisotropy. The absorption curves for the $xx$ and $yy$ polarizations exhibit very similar line shapes throughout the investigated energy range, differing only slightly in peak intensity and position. This behavior indicates that the dipole transition matrix elements are nearly isotropic within the monolayer plane despite the reduced crystal symmetry.

\subsection{Topological Properties}

\begin{figure}[H]
\centering
\begin{subfigure}{0.45\textwidth}
\includegraphics[width=\textwidth,clip=true]{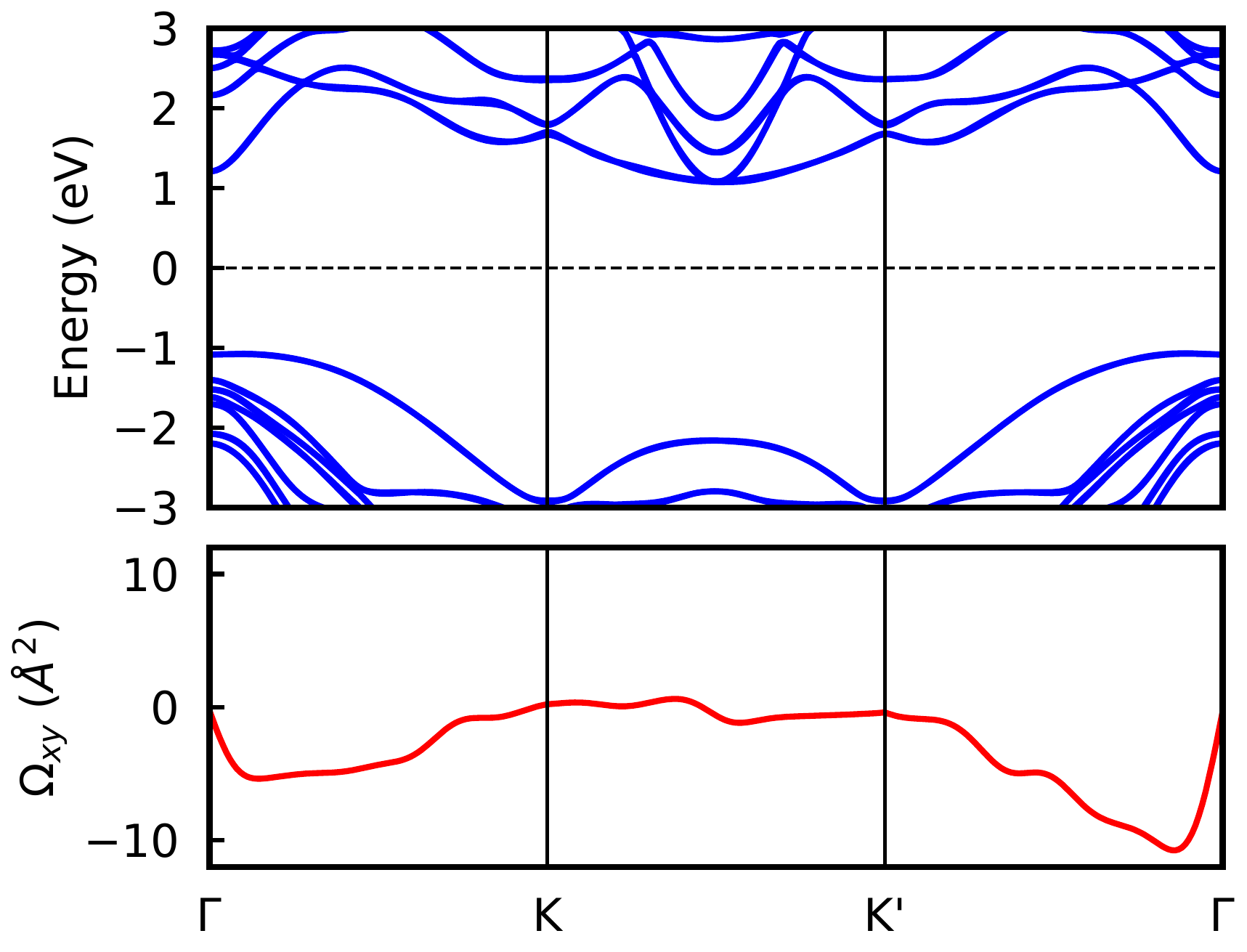}
\caption{}
\label{fig:first}
\end{subfigure}
\hfill
\begin{subfigure}{0.45\textwidth}
\includegraphics[width=\textwidth,clip=true]{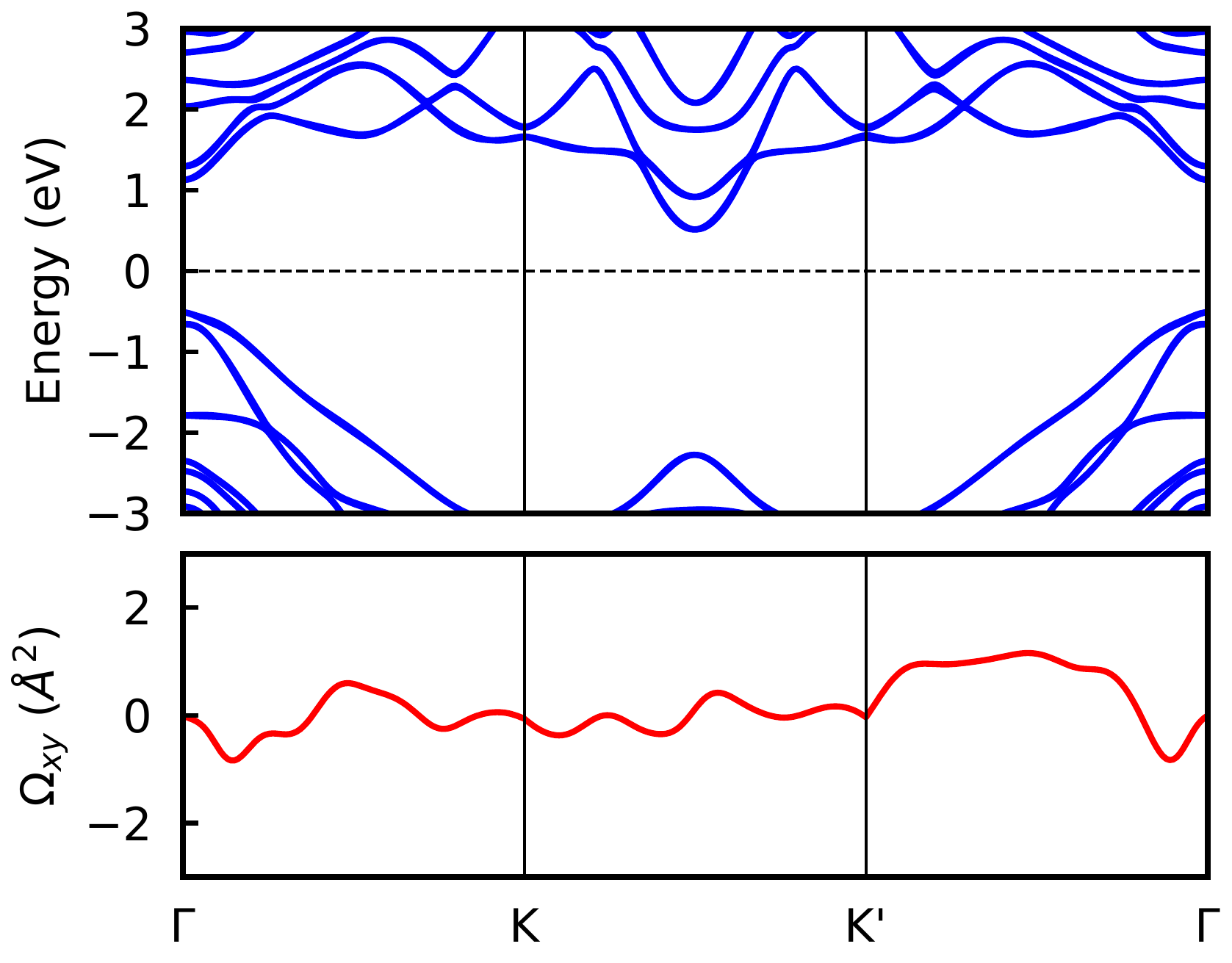}
\caption{}
\label{fig:second}
\end{subfigure}
\hfill
\begin{subfigure}{0.45\textwidth}
\includegraphics[width=\textwidth,clip=true]{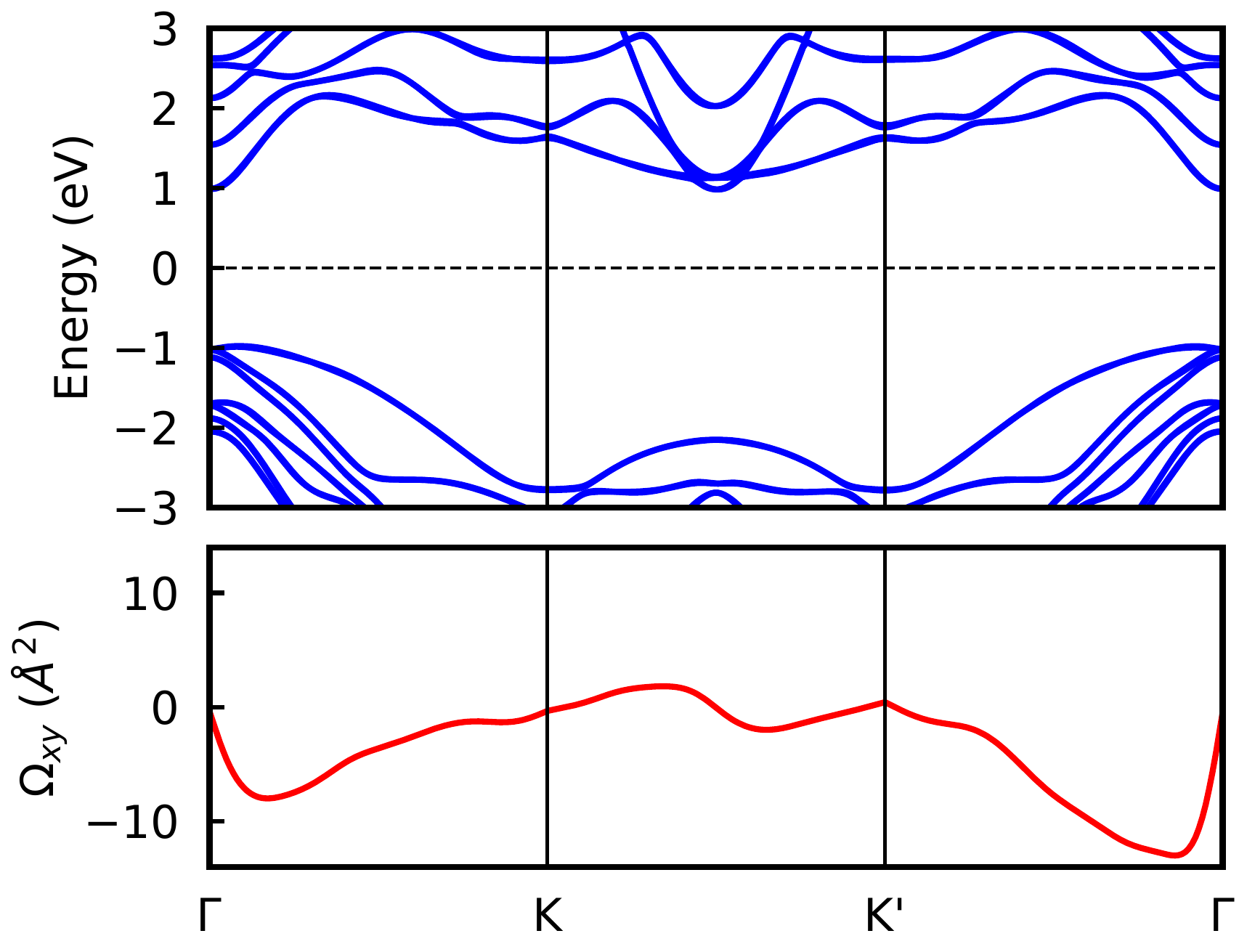}
\caption{}
\label{fig:third}
\end{subfigure}
\hfill
\begin{subfigure}{0.45\textwidth}
\includegraphics[width=\textwidth,clip=true]{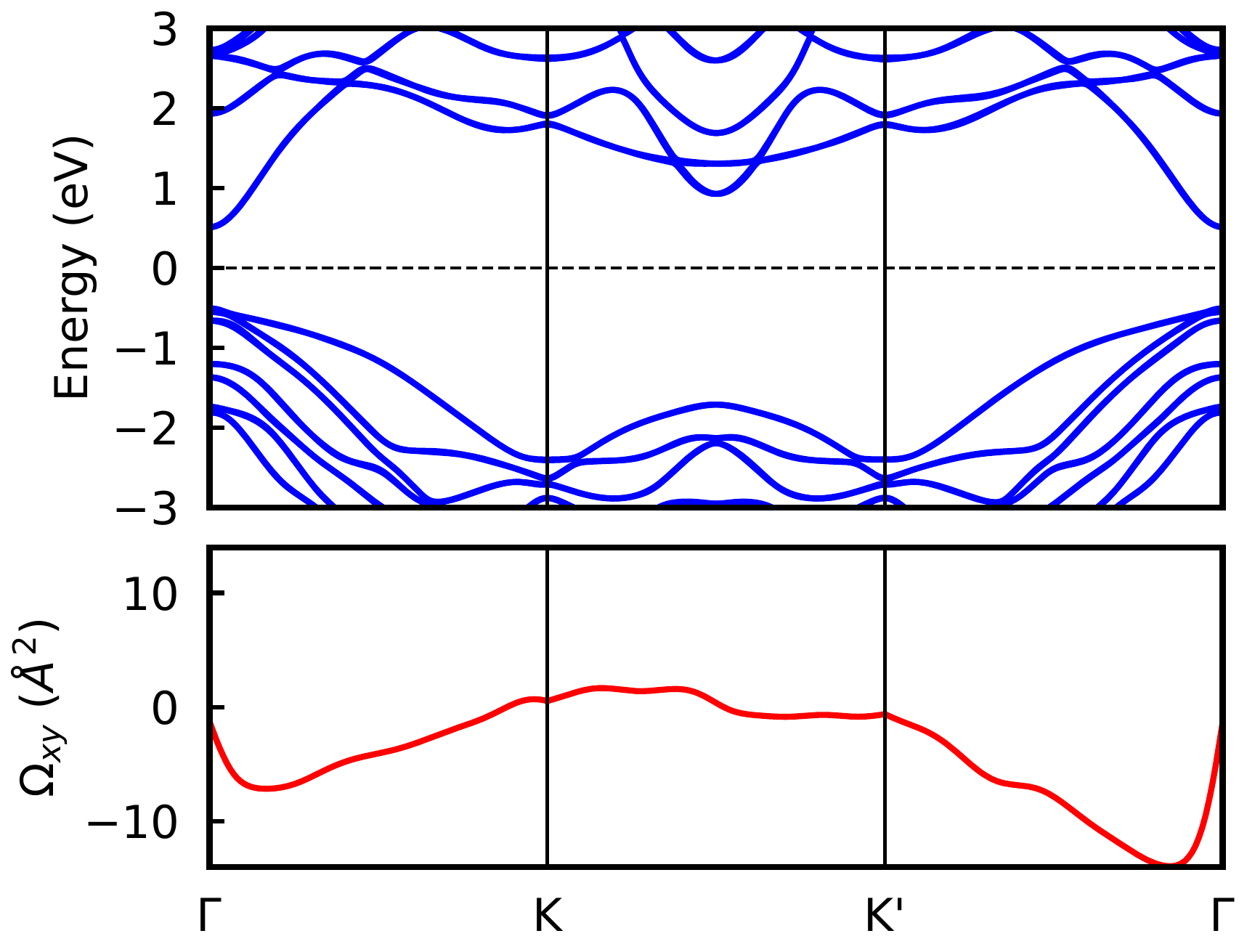}
\caption{}
\label{fig:berry}
\end{subfigure}
\caption{Electronic band structures and corresponding Berry curvature $\Omega_{xy}(\mathbf{k})$ evaluated along the high-symmetry path $\Gamma$--K--K$'$--$\Gamma$ for (a) \ce{Si2Al2Te2}, (b) \ce{Si2B2Te2}, (c) \ce{Si2Ga2Te2}, and (d) \ce{Si2In2Te2}.}
\end{figure}

\begin{figure}[htbp]
    \centering
    \includegraphics[width=0.5\linewidth]{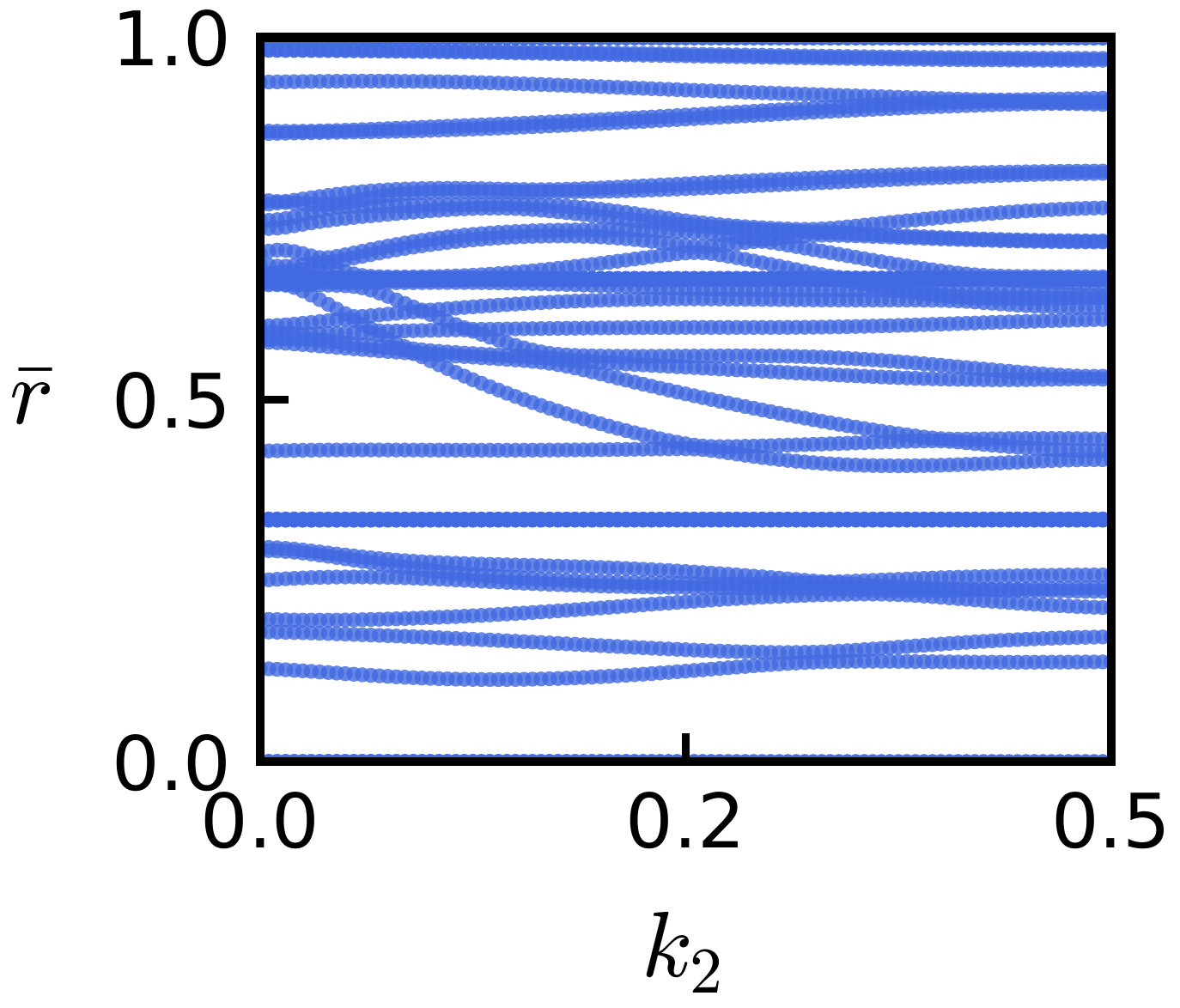}
    \caption{Evolution of the Wannier charge centers (WCCs) for the
$\mathrm{Si_2In_2Te_2}$ monolayer calculated at the HSE06+SOC level
on the physical $k_3=0$ plane. The nontrivial evolution of the WCCs,
characterized by partner switching, yields the two-dimensional
topological invariant $Z_2=1$.}
    \label{fig:wcc}
\end{figure}

The topological character of $\mathrm{Si_2In_2Te_2}$ was further
examined through the evolution of the Wannier charge centers (WCCs),
shown in Fig.~\ref{fig:wcc}. Since the system is a two-dimensional
monolayer, the relevant topological classification is given by a
single two-dimensional $Z_2$ invariant. We therefore consider the
physical $k_3=0$ plane of the Brillouin zone.

For each fixed value of $k_2$, the Wilson loop is evaluated along the
closed reciprocal-space direction $k_1$. The eigenphases of the
Wilson-loop operator define the hybrid Wannier charge centers
$\bar{r}_n(k_2)$, whose evolution is followed over half of the
time-reversal-invariant Brillouin zone, from $k_2=0$ to $0.5$.

As shown in Fig.~\ref{fig:wcc}, the WCC branches exhibit partner
switching as $k_2$ evolves between the two time-reversal-invariant
boundaries. Equivalently, a reference line crossing the WCC flow
intersects the branches an odd number of times. This odd connectivity
yields a nontrivial two-dimensional invariant, $Z_2=1$, 
demonstrating that $\mathrm{Si_2In_2Te_2}$ is a two-dimensional
quantum spin Hall insulator. This result provides the topological
classification of the insulating state independently of the detailed
orbital interpretation of the SOC-induced changes in the band
structure.

The Berry curvature $\Omega_{xy}(\mathbf{k})$ evaluated along the
high-symmetry path $\Gamma$--K--K$'$--$\Gamma$ is shown in
Fig.~\ref{fig:berry}. The Berry curvature exhibits pronounced
momentum-dependent features in the vicinity of closely spaced and
avoided-crossing bands, reflecting the strong interband coupling
present in these regions. As required by time-reversal symmetry,
$\Omega_{xy}(\mathbf{k})=-\Omega_{xy}(-\mathbf{k})$, so that the
Berry curvature integrated over the full Brillouin zone vanishes.
Consequently, the local Berry-curvature features should not by
themselves be interpreted as evidence of a nontrivial $Z_2$
topology. The topological classification is instead determined from
the Wilson-loop/Wannier-charge-center analysis discussed below.

The calculated $\mathbb{Z}_2$ topological invariants classify \ce{Si2B2Te2}, \ce{Si2Al2Te2}, and \ce{Si2Ga2Te2} as topologically trivial ($\mathbb{Z}_2 = 0$). In contrast, \ce{Si2In2Te2} exhibits a nontrivial topological invariant ($\mathbb{Z}_2 = 1$). As discussed above, the
In-containing compound exhibits a more pronounced SOC-induced
reorganization of the strongly hybridized $p$-derived states near
the band edges. The nontrivial character of the resulting insulating
state is established by the evolution of the Wannier charge centers,
as discussed below.

\section{Conclusions}

In summary, we have performed a comprehensive first-principles investigation of the structural, vibrational, electronic, optical, and topological properties of two-dimensional \ce{Si2X2Te2} monolayers (\ce{X}= \ce{B}, \ce{Al}, \ce{Ga}, and \ce{In}). Phonon dispersion calculations confirm the harmonic dynamical stability of all four systems, with vibrational frequencies scaling inversely with the atomic mass of the group-III element. Electronic structure calculations demonstrate that all monolayers are semiconductors whose valence-band edges are primarily composed of \ce{Te} $p$ states, while the conduction-band edges exhibit a hybridized \ce{Si}--\ce{X}--\ce{Te} character. 

Finally, the topological analysis demonstrates that chemical
substitution can tune the band topology of the
$\mathrm{Si_2X_2Te_2}$ family. While the B-, Al-, and Ga-based
systems are topologically trivial ($Z_2=0$),
$\mathrm{Si_2In_2Te_2}$ exhibits a stronger SOC-induced
reorganization of its hybridized $p$-derived electronic states and
realizes a nontrivial quantum spin Hall phase ($Z_2=1$).

\begin{acknowledgement}

The authors also express their gratitude to Santos Dumont/LNCC, LAMCAD/UFG, CENAPAD/SP and Lobo Carneiro HPC (NACAD) at the Federal University of Rio de Janeiro (UFRJ) for computer resources. A.C.D. acknowledges financial support from FAP-DF grant numbers 00193-00001817/2023-43 and 00193-00002073/2023-84, CNPq grant numbers 408144/2022-0, 305174/2023-1, 444069/2024-0, and 444431/2024-1. A.C.D and A.L.R also acknowledge funding from PDPG-FAPDF-CAPES Centro-Oeste grant number 00193-00000867/2024-94 and CAPES-COFECUB grant number 88881.188740/2025-01.   
 
\end{acknowledgement}

\bibliography{refs}

\clearpage
\newpage

\renewcommand{\thepage}{S-\arabic{page}}
\renewcommand{\thefigure}{S\arabic{figure}}
\renewcommand{\thetable}{S\arabic{table}}
\renewcommand{\thesection}{S\arabic{section}}

{\bf \Large {Supporting Information}}

\section{PAW Projectors: Computational Technical Details}

\begin{table}[H]
\centering\caption{All POTCAR files were obtained from the potpaw\_pbe\_6.4 library with the GW variant. Most important information of the selected PAW projectors, which includes PAW-PBE projector name, date of projector creation, number of valence electrons, Zval and maximum recommended cutoff energy, ENMAX, for all selected chemical species.}
\begin{tabular}{clccc} \toprule
Element & \texttt{POTCAR} & Date & $Z_{val}$    & \texttt{ENMAX} \\
        & PAW-PBE & \texttt{POTCAR}    &      & (\si{\electronvolt}) \\ \midrule
\ce{Si} &\ce{Si}\_GW    & 05/04/2012  & 4   & 245.345  \\
\ce{B} &\ce{B}\_GW\_new & 03/26/2016  & 3   & 318.614  \\
\ce{Ga} &\ce{Ga}\_GW    & 03/22/2012  & 3   & 134.678  \\
\ce{In} &\ce{In}\_d\_GW & 03/15/2013  & 13  & 278.624  \\
\ce{Al} &\ce{Al}\_GW    & 03/19/2012  & 3   & 240.300  \\
\ce{Te} &\ce{Te}\_GW    & 03/22/2012  & 6   & 174.982   \\
\bottomrule
\end{tabular}
\end{table}

\section{Crystal Structures in POSCAR format}

\subsection*{\ce{Si2B2Te2}}

\texttt{Si2B2Te2  \\                                 
   1.0000000000000000     \\
     3.6783816648842516   -0.0002999709188704    0.0000000000000000 \\
    -1.8389941622817543    3.1858208531473275    0.0000000000000000 \\
     0.0000000000000000    0.0000000000000000   28.1709999999999994 \\
   Si   B    Te \\
     2     2     2 \\
Direct \\
     0.9999311815110161    0.0000256465818822    0.6933624129061684 \\
     0.3332753153336654    0.6667034919158539    0.6616124128455709 \\
     0.9999175461738830    0.0000120136770647    0.7647146372364375 \\
     0.3332886806000346    0.6667131775864661    0.5902600218597485 \\
     0.6666104717575010    0.3333686277309056    0.5548053458547955 \\
     0.6665954128211808    0.3333557269204661    0.8001708836844230}

\subsection*{\ce{Si2Al2Te2}}

\texttt{Si2Al2Te2 \\                                
   1.0000000000000000 \\     
     4.0219624501734224   -0.0003430266801598    0.0000202561407786 \\
    -2.0107782562106586    3.4834863959651643    0.0000131998498130 \\
     0.0000000000000000    0.0000000000000000   28.1709999999999994 \\
   Si   Al   Te \\
     2     2     2 \\
Direct \\
     0.9999357985346862    0.0000373659674224    0.6905769734187928 \\
     0.3332640407811240    0.6666996409200578    0.6643972641578983 \\
     0.9998948331633315    0.0000019647151390    0.7774531869947907 \\
     0.3332937800514628    0.6667339459887245    0.5775204205220561 \\
     0.6665919865796255    0.3333638433634363    0.5309964098451232 \\
     0.6665980625751544    0.3333709158596747    0.8239761138803416}

\subsection*{\ce{Si2Ga2Te2}}

\texttt{Si2Ga2Te2   \\                                
   1.0000000000000000  \\    
     4.0257962177298587   -0.0002688941390919    0.0000000000000000 \\
    -2.0126309742580442    3.4864869194180836    0.0000000000000000 \\
     0.0000000000000000    0.0000000000000000   28.1709999999999994 \\
   Si   Ga   Te \\
     2     2     2 \\
Direct \\
     0.9999365138295460    0.0000356294250494    0.6903023117540883 \\
     0.3332706241215533    0.6667034211326595    0.6646570473919766 \\
     0.9999363582353809    0.0000354264056170    0.7765466845107696 \\
     0.3332611305760125    0.6666899794014256    0.5784135957391072 \\
     0.6666140471989408    0.3333803717696497    0.5316011911219860 \\
     0.6665835411796763    0.3333449118898102    0.8233600469802198}

\subsection*{\ce{Si2In2Te2}}

\texttt{Si2In2Te2   \\                                
   1.0000000000000000  \\    
     4.1513320759845236   -0.0002679150760732    0.0000000000000000 \\
    -2.0753826953385803    3.5950889011773981    0.0000000000000000 \\
     0.0000000000000000    0.0000000000000000   28.1709999999999994 \\
   Si   In   Te \\
     2     2     2 \\
Direct \\
     0.9999271657417808    0.0000343051216518    0.6899463009793436 \\
     0.3332673008182709    0.6667087738368025    0.6650347177519791 \\
     0.9999292057390434    0.0000263631729780    0.7826351683326962 \\
     0.3332758148775028    0.6666941032493625    0.5723423765319708 \\
     0.6666435022928141    0.3334122264110064    0.5191187465036151 \\
     0.6665612546288173    0.3333083826658552    0.8358612078798942}

\section{Bands}

\begin{figure}[H]
\centering
\begin{subfigure}{0.68\textwidth}
\subcaption[]{}
\includegraphics[width=\textwidth,clip=true]{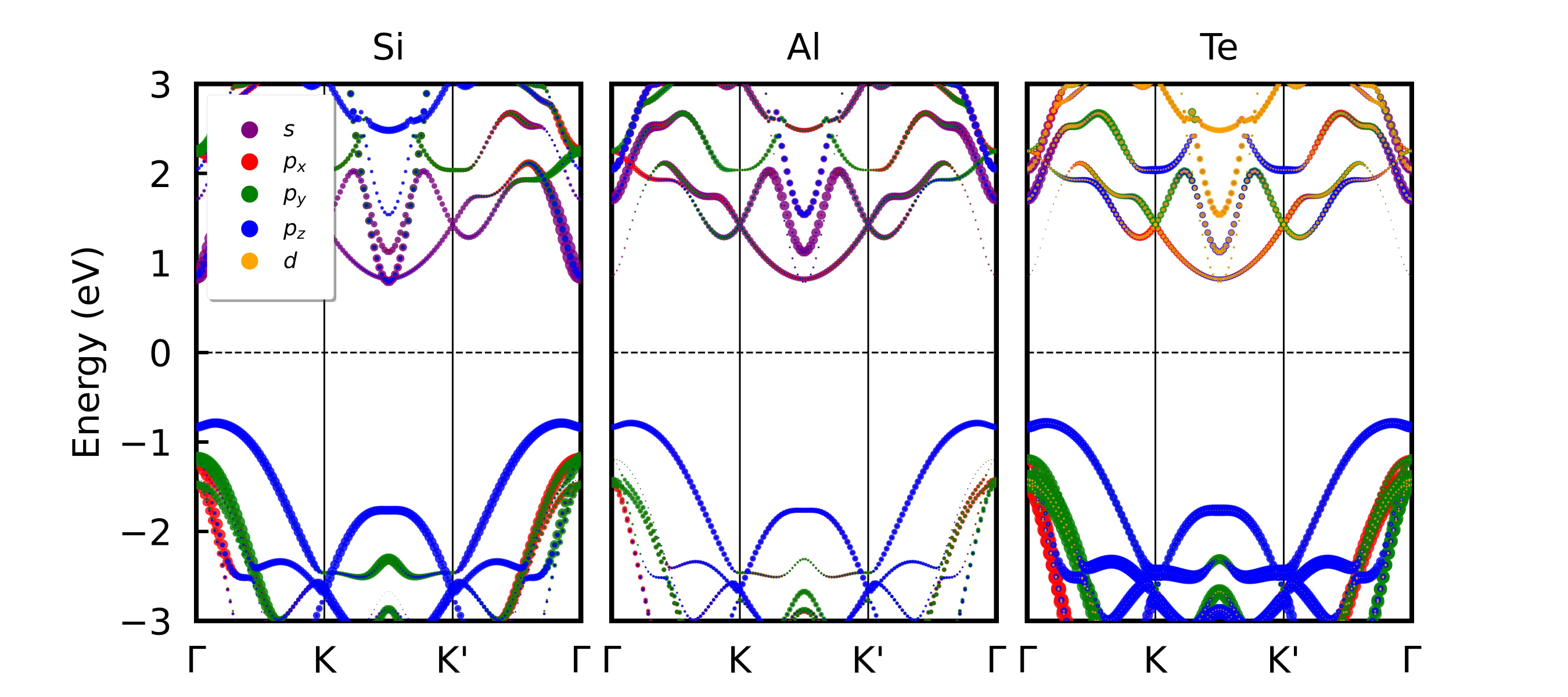}
\end{subfigure}
\hfill
\begin{subfigure}{0.68\textwidth}
\subcaption[]{}
\includegraphics[width=\textwidth,clip=true]{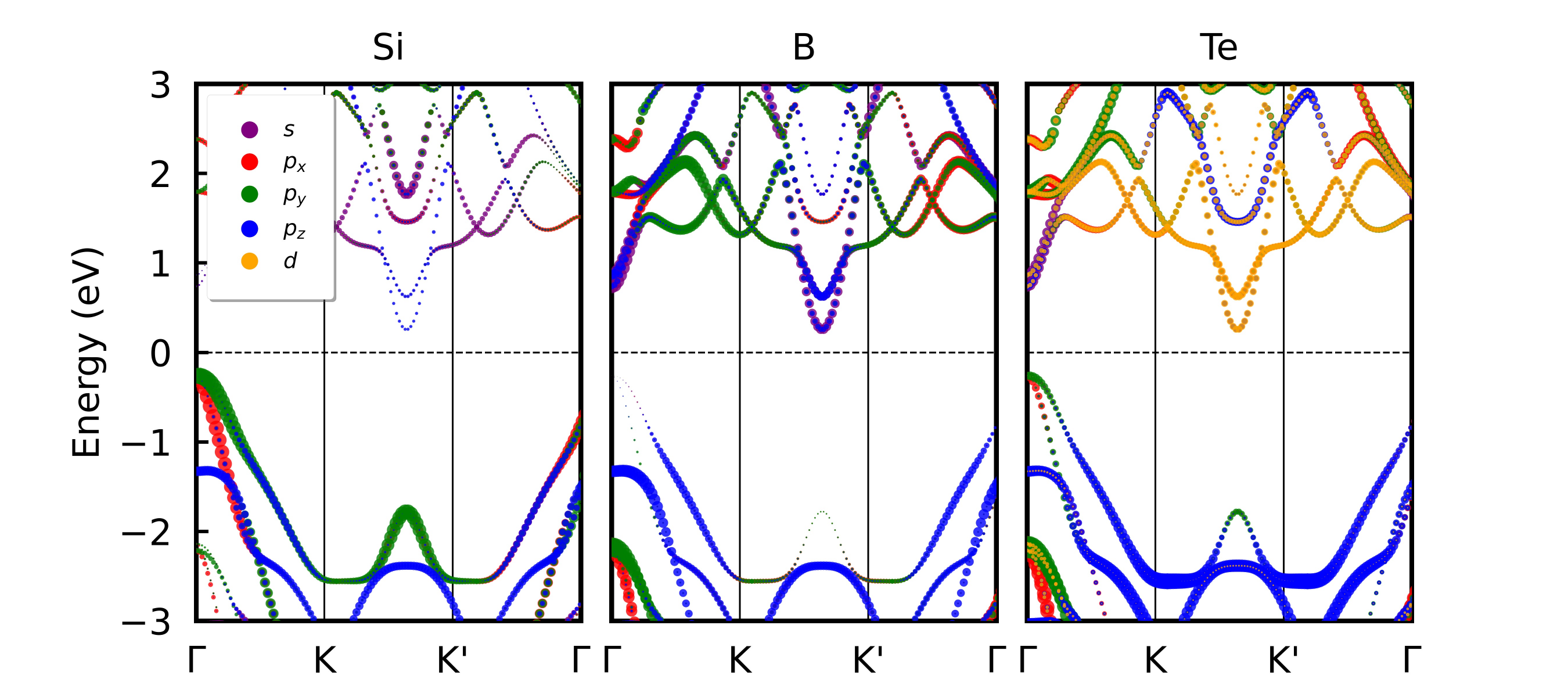}
\end{subfigure}
\hfill
\begin{subfigure}{0.68\textwidth}
\subcaption[]{}
\includegraphics[width=\textwidth,clip=true]{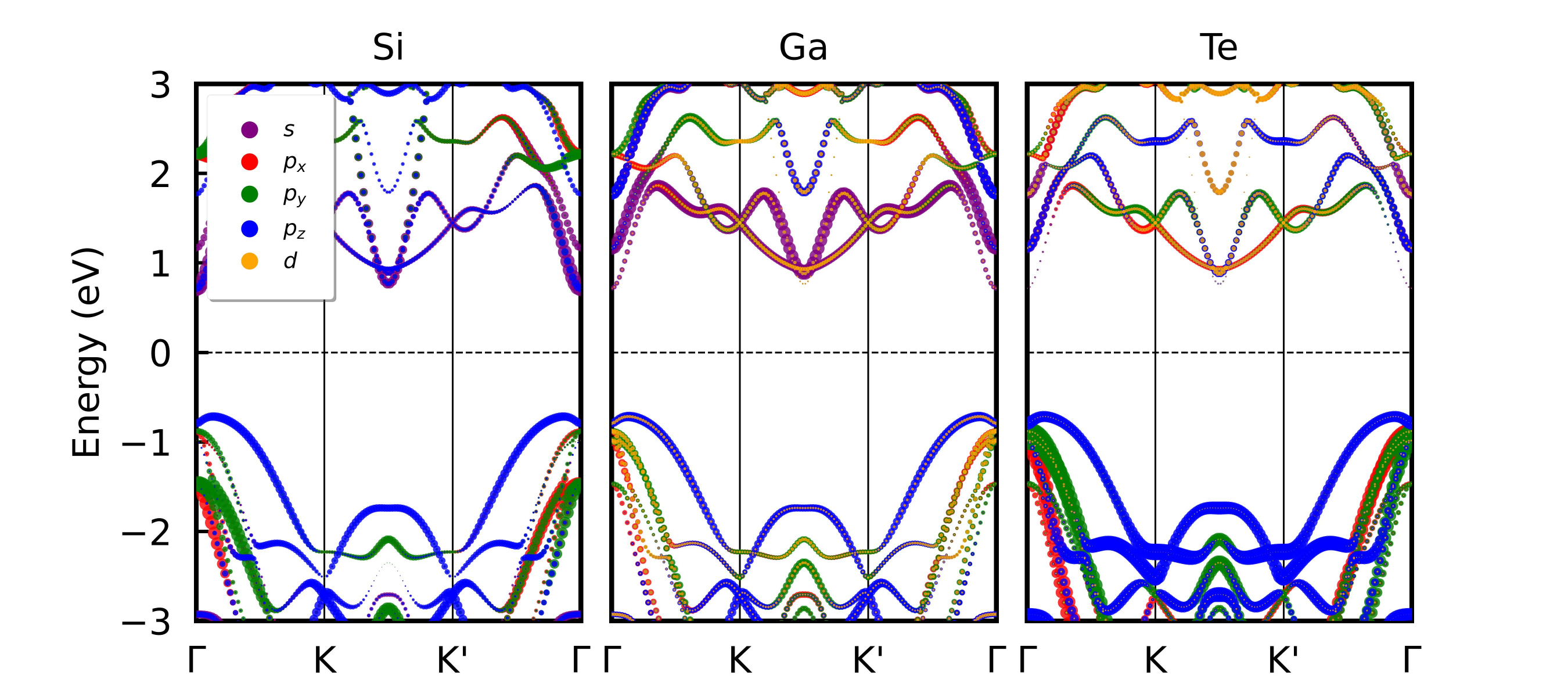}
\end{subfigure}
\hfill
\begin{subfigure}{0.68\textwidth}
\subcaption[]{}
\includegraphics[width=\textwidth,clip=true]{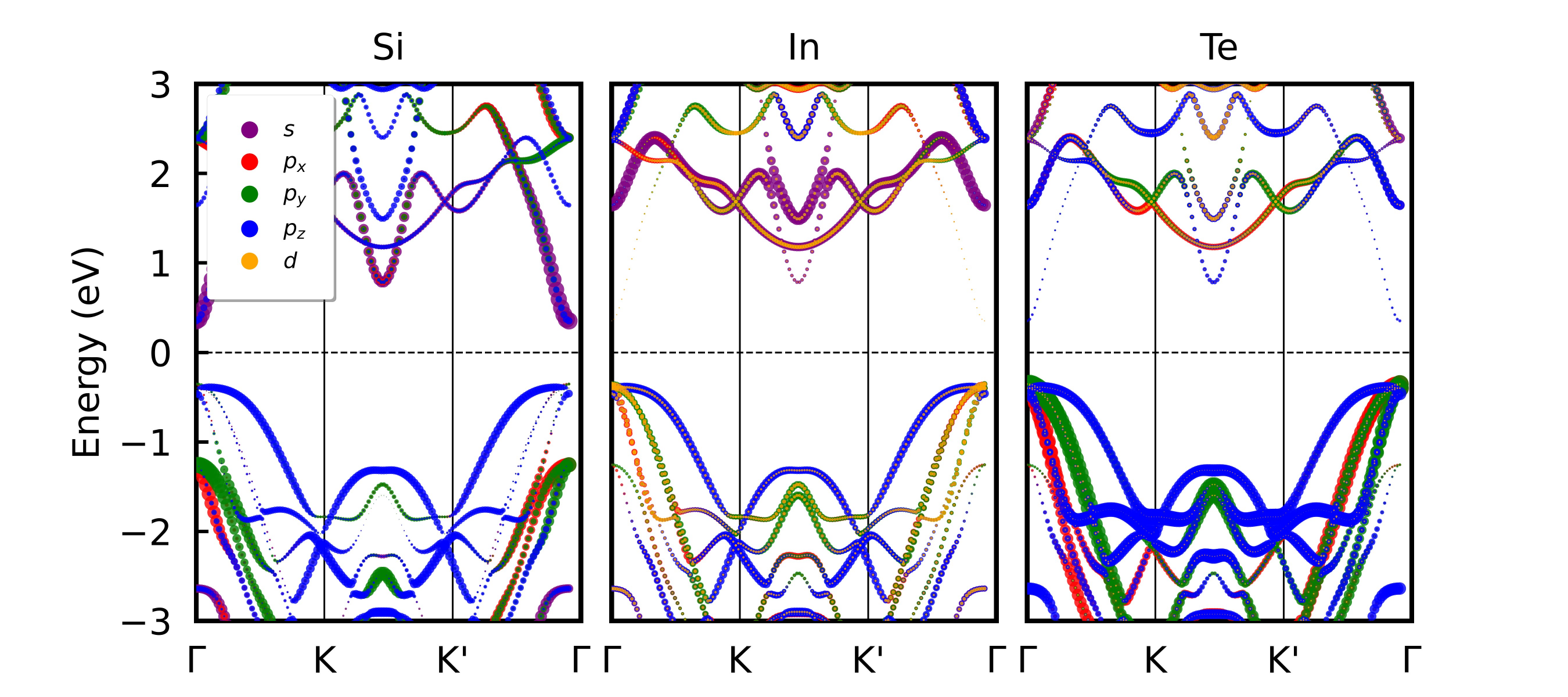}
\end{subfigure}
\caption{\label{fig:bands_proj_hse} Electronic band structure calculated with PBE without SOC of (a) \ce{Si2Al2Te2}, (b) \ce{Si2B2Te2}, (c) \ce{Si2Ga2Te2}, and (d) \ce{Si2In2Te2}.}
\end{figure}

\begin{figure}[H]
\centering
\begin{subfigure}{0.68\textwidth}
\subcaption[]{}
\includegraphics[width=\textwidth,clip=true]{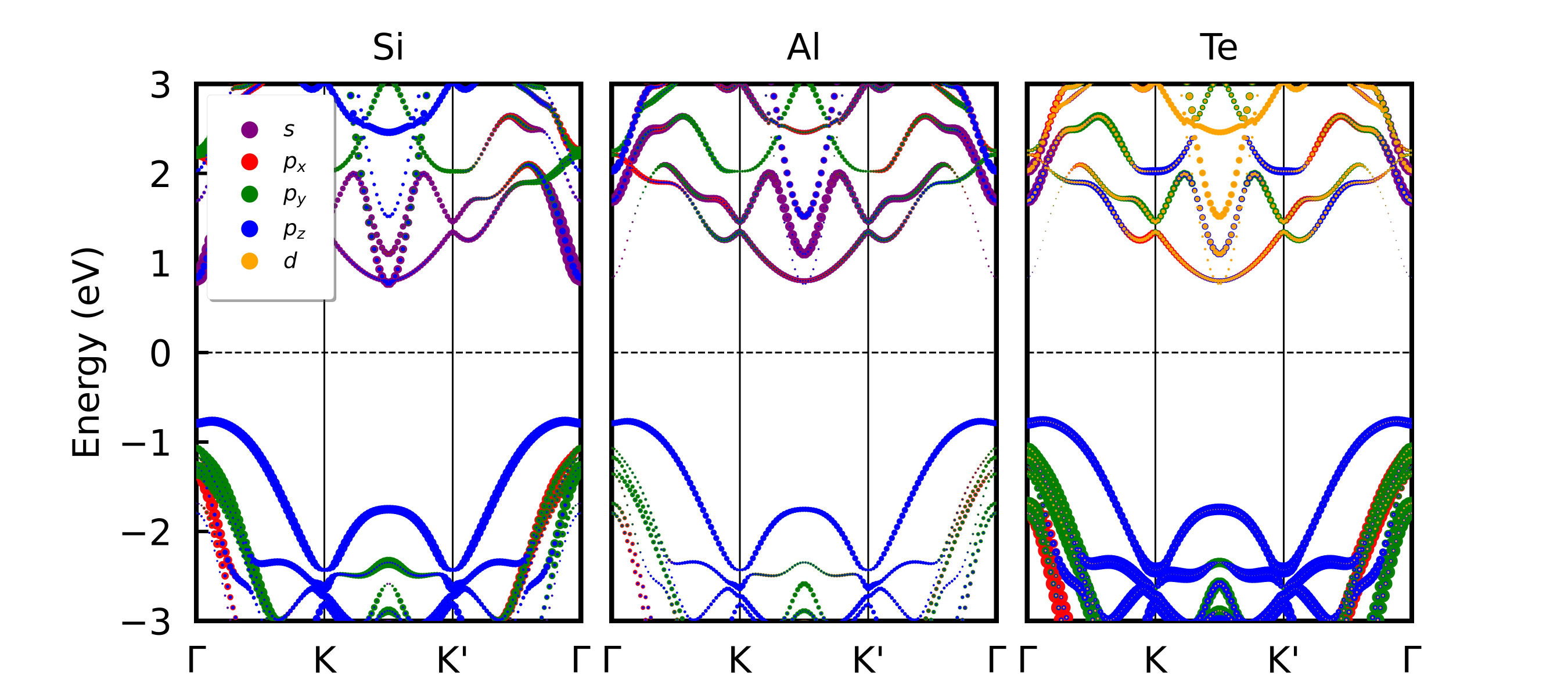}
\end{subfigure}
\hfill
\begin{subfigure}{0.68\textwidth}
\subcaption[]{}
\includegraphics[width=\textwidth,clip=true]{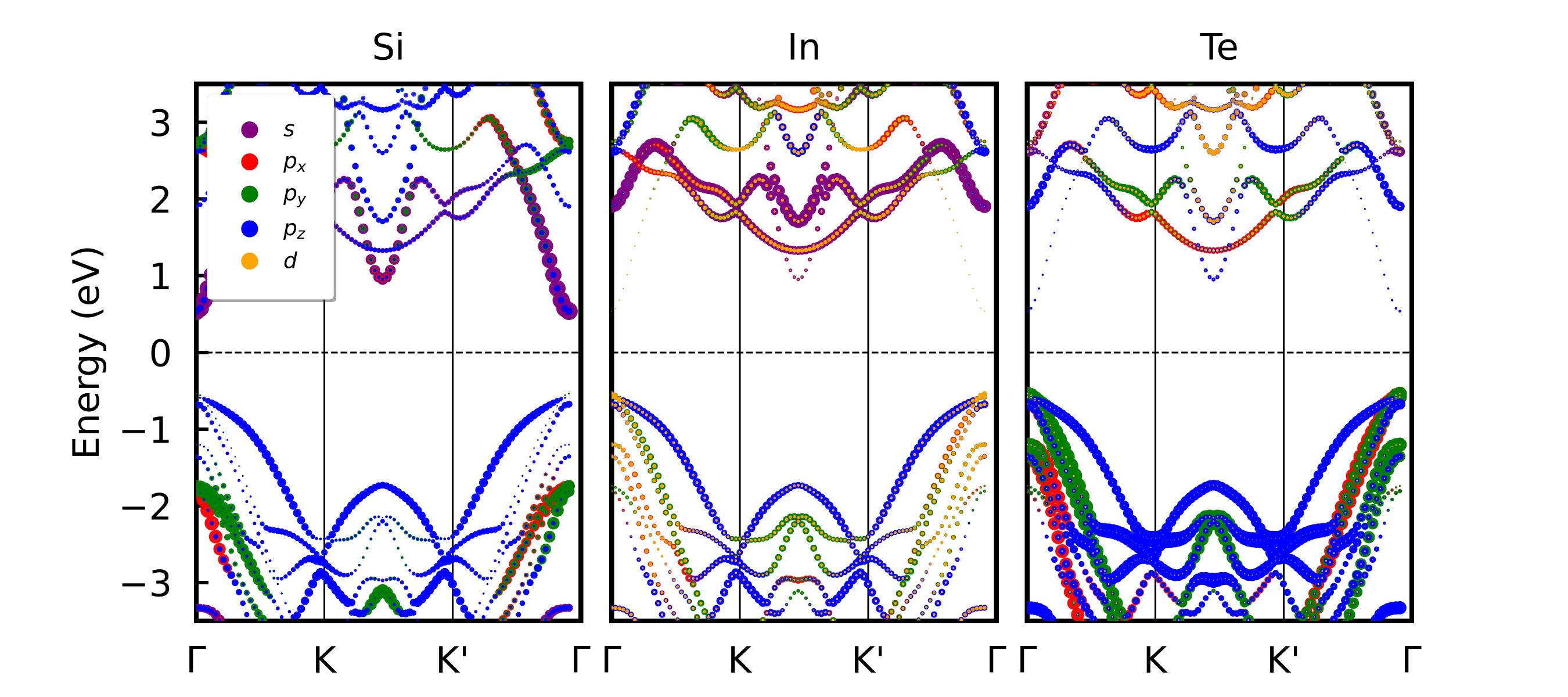}
\end{subfigure}
\hfill
\begin{subfigure}{0.68\textwidth}
\subcaption[]{}
\includegraphics[width=\textwidth,clip=true]{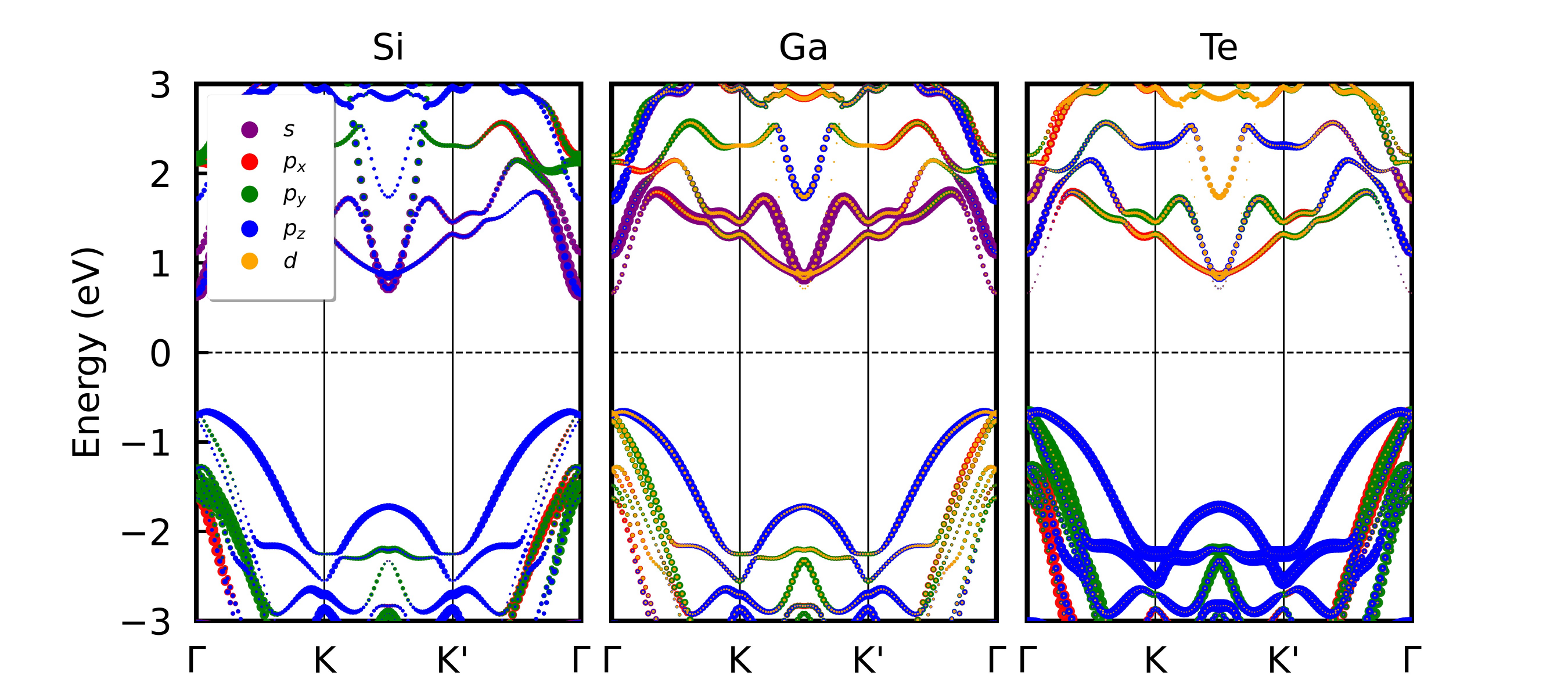}
\end{subfigure}
\hfill
\begin{subfigure}{0.68\textwidth}
\subcaption[]{}
\includegraphics[width=\textwidth,clip=true]{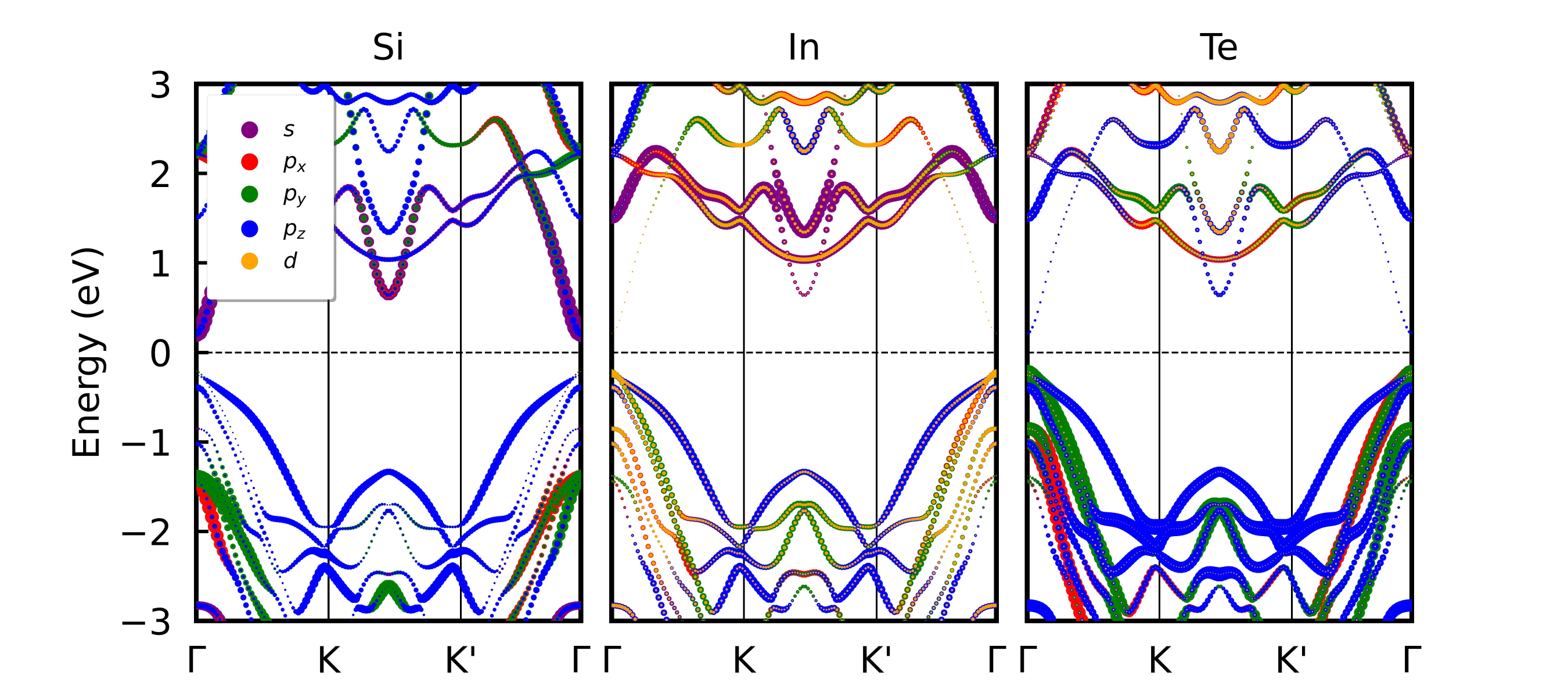}
\end{subfigure}
\caption{\label{fig:bands_proj_hse} Electronic band structure calculated with PBE with SOC of (a) \ce{Si2Al2Te2}, (b) \ce{Si2B2Te2}, (c) \ce{Si2Ga2Te2}, and (d) \ce{Si2In2Te2}.}
\end{figure}

\begin{figure}[!hbt]
\centering
\includegraphics[width=1\linewidth,clip=true]{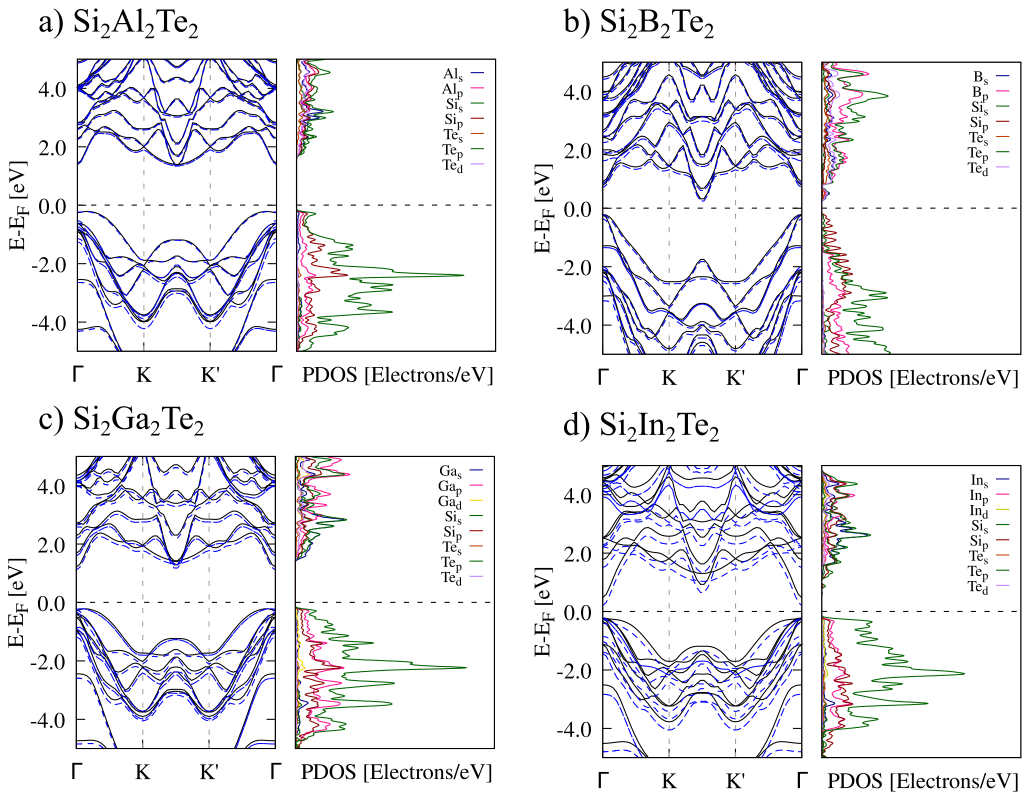}
\caption{Electronic band structure and Density of States (DOS). The left panel shows the calculated band dispersion, where the solid black and dashed blue lines represent the PBE and PBE + SOC calculations, respectively. The right panel displays the corresponding DOS.}
\label{fig:bands-pbe}
\end{figure}

\begin{figure}[!hbt]
\centering
\includegraphics[width=1\linewidth,clip=true]{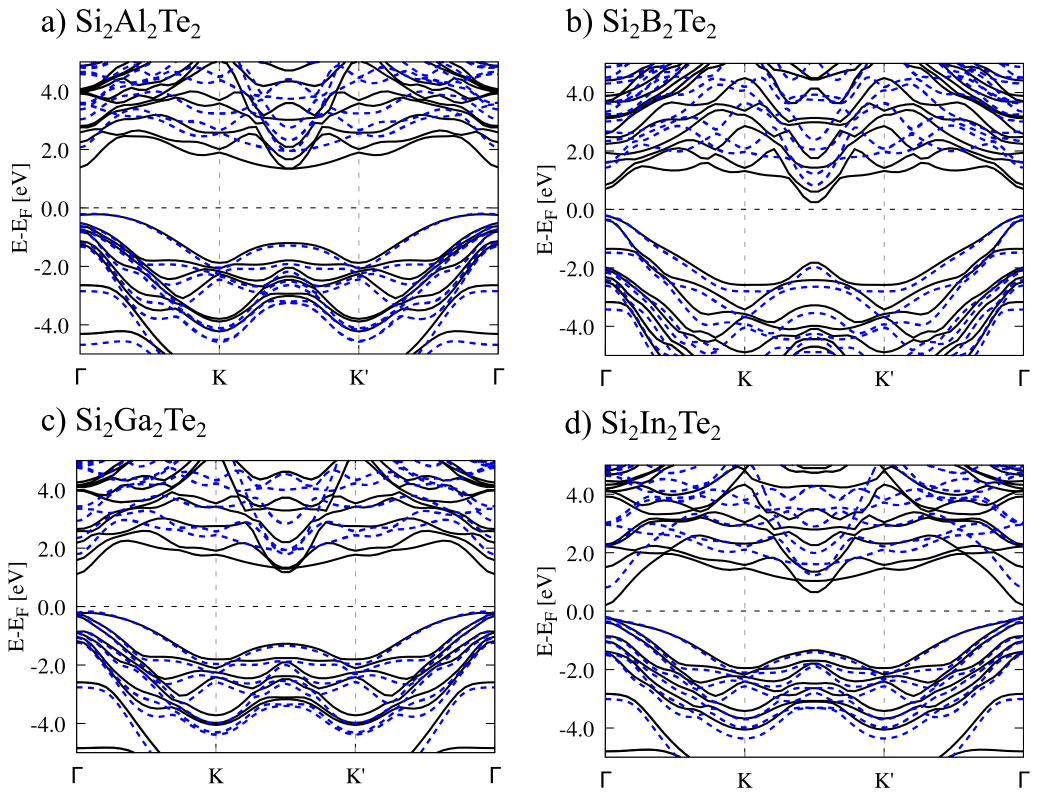}
\caption{Electronic band structure. The solid black and dashed blue lines represent the calculations performed with PBE+SOC and HSE06+SOC, respectively.}
\label{fig:bands-hse}
\end{figure}

\clearpage
\newpage

\section{BSE Simulation parameters}

\begin{table}[H]
\centering
\caption{Parameters used for BSE simulations:\textbf{k}-points density, $R_{k}$ (\si{\per\angstrom}) and their correspondent \textbf{k}-mesh, $n_v$, number of valence bands, $n_c$, number of conduction bands and dielectric function smearing $\eta$ (\si{\electronvolt}). All simulations were done using the Coulomb truncated 2D potential (V2DT) with the system surrounded by vacuum, implemented in WanTiBEXOS package.}

    \begin{tabular}{lcccccc} \toprule
    Structure  & $R_k$ & \textbf{k}-mesh   & $n_c$ & $n_v$ & $\eta$ \\ \midrule

    Si$_2$Al$_2$Te$_2$ &120  &$34\times34\times1$ &10 &10   & 0.05\\
    Si$_2$B$_2$Te$_2$ &120  &$38\times38\times1$  &9  &9  & 0.05\\
    Si$_2$Ga$_2$Te$_2$ &120  &$34\times34\times1$ &9  &12  & 0.05\\  
    Si$_2$In$_2$Te$_2$ &120  &$33\times33\times1$ &10 &12   & 0.05\\
     \bottomrule
    \end{tabular}
    \label{tab:bse_params}
\end{table}

\begin{figure}[!hbt]
    \centering
    \includegraphics[width=1\linewidth,clip=true]{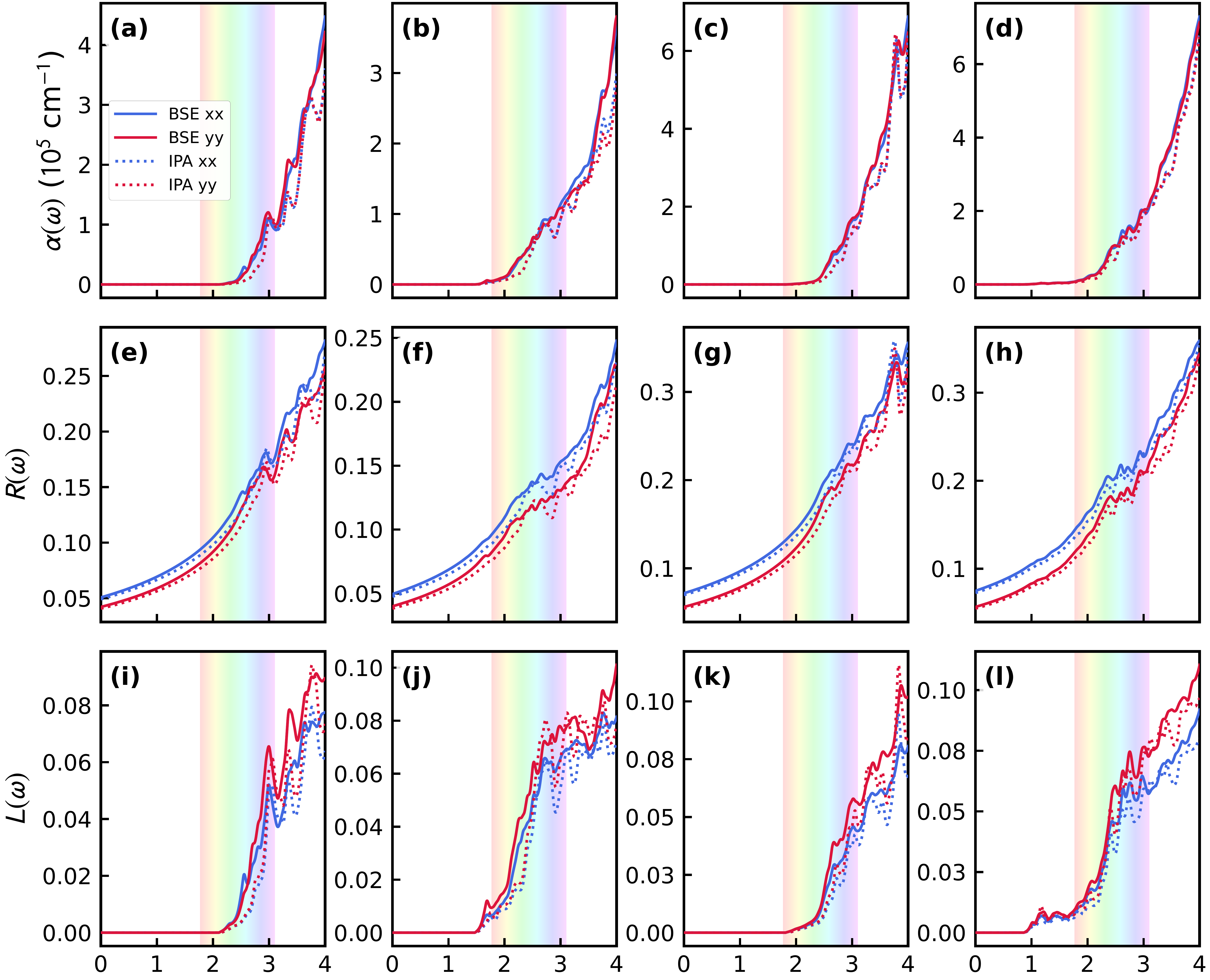}
    \caption{Calculated optical absorption coefficient ($\alpha(\omega)$), reflectivity ($R(\omega)$), and electron energy-loss function ($L(\omega)$) for (a-d) Si$_2$Al$_2$Te$_2$, (e-h) Si$_2$B$_2$Te$_2$, (i-l) Si$_2$Ga$_2$Te$_2$, and (m-p) Si$_2$In$_2$Te$_2$. The optical spectra were obtained within the independent-particle approximation (IPA, dashed lines) and by solving the Bethe–Salpeter equation (BSE, solid lines). Blue and red curves correspond to light polarized along the crystallographic $x$ and $y$ directions, respectively. The shaded region indicates the visible spectral range.}\label{fig:bands-hse}
\end{figure}

\section{Berry phase and Berry curvature}

The topological properties of the two-dimensional phases were characterized using the Chern number and the  $\mathbb{Z}_2$  invariant,
both evaluated from the Berry curvature and the evolution of the
occupied electronic subspace in momentum space. For systems that break time-reversal symmetry, the Chern number $C$ is
defined as the Brillouin-zone integral of the Berry curvature summed
over all occupied bands,

\begin{equation}
C = \frac{1}{2\pi} \int_{\mathrm{BZ}} \Omega_{xy}(\mathbf{k}) \, d^2k ,
\end{equation}

where $\Omega_{xy}(\mathbf{k})$ is the Berry curvature summed over the occupied
bands, $\Omega_{xy}(\mathbf{k})=\sum_{n\in\mathrm{occ}}\Omega_{n,xy}(\mathbf{k})$.
In our implementation, $\Omega_{n,xy}(\mathbf{k})$ is evaluated using the
Kubo (velocity-matrix) expression, as implemented in \textsc{WannierTools},

  \begin{equation}
\Omega_{n,xy}(\mathbf{k}) =
-2\,\mathrm{Im}\sum_{m\neq n}
\frac{
\langle u_{n\mathbf{k}}|\,\hat v_x\,|u_{m\mathbf{k}}\rangle\,
\langle u_{m\mathbf{k}}|\,\hat v_y\,|u_{n\mathbf{k}}\rangle
}{
\left(\varepsilon_{m\mathbf{k}}-\varepsilon_{n\mathbf{k}}\right)^2
},
\end{equation}
where $|u_{n\mathbf{k}}\rangle$ and $\varepsilon_{n\mathbf{k}}$ are the
cell-periodic eigenstates and eigenvalues of the Wannier-interpolated tight-binding
Hamiltonian $H(\mathbf{k})$, and the velocity operators are obtained from
$\hat v_{\alpha}=(1/\hbar)\,\partial H(\mathbf{k})/\partial k_{\alpha}$.
In time-reversal-symmetric systems, the Berry curvature satisfies
$\Omega_{xy}(\mathbf{k})=-\Omega_{xy}(-\mathbf{k})$, enforcing a vanishing total
Chern number, although sizable local Berry-curvature contributions may still occur.
For time-reversal-invariant two-dimensional systems, the relevant topological index
is the $\mathbb{Z}_2$ invariant, which distinguishes trivial insulators from quantum
spin Hall phases. The $\mathbb{Z}_2$ invariant was determined from the evolution of
the Wilson loop, or equivalently the Wannier charge centers, of the occupied bands
across the Brillouin zone. A nontrivial topological phase corresponds to an odd
winding of the Wannier charge centers and yields $\mathbb{Z}_2=1$, whereas the
absence of winding indicates a trivial phase with $\mathbb{Z}_2=0$.

The Berry curvature and Chern number were computed using tight-binding
Hamiltonians constructed from MLWF ensuring an accurate interpolation
of the electronic structure on dense $\mathbf{k}$-point
meshes.

The topological behavior of the 2D materials is shown to be highly
sensitive to lattice geometry, buckling, and symmetry breaking,
leading to a hierarchy of quantum phases within the same chemical
composition. For time-reversal-invariant two-dimensional systems, the relevant
topological index is the $Z_2$ invariant, which distinguishes trivial
insulators from quantum spin Hall phases. The $Z_2$ invariant was
determined from the evolution of the Wilson loop, equivalently the
Wannier charge centers (WCCs), of the occupied electronic manifold.

A nontrivial evolution of the WCCs characterized by partner switching
yields $Z_2=1$, whereas a trivial WCC connectivity yields $Z_2=0$.
Spin--orbit coupling was included in all topological calculations.

Within the $\mathrm{Si_2X_2Te_2}$ series, the B-, Al-, and Ga-containing
monolayers are topologically trivial ($Z_2=0$), whereas
$\mathrm{Si_2In_2Te_2}$ exhibits a nontrivial $Z_2=1$ phase.

\end{document}